\documentclass[a4paper,fleqn,usenatbib]{mnras}
\pdfoutput=1

\usepackage[T1]{fontenc}
\usepackage{ae,aecompl}

\usepackage{graphicx}	        
\graphicspath{{figures/}}
\usepackage{amsmath}           	
\usepackage{amssymb}	        

\usepackage{siunitx}            
\usepackage{booktabs}           
\usepackage{threeparttable}     

\usepackage[font=footnotesize, caption=false]{subfig}

\usepackage[final,english,marginclue,footnote]{fixme}
\FXRegisterAuthor{ref}{anref}{\textsf{Ref.}}

\fxusetheme{color}

\usepackage{xspace}		
\usepackage{bm}         

\usepackage{algorithm}      
\usepackage{algorithmic}    

\renewcommand{\vec}[1]{\bm{\mathrm{#1}}}

\newcommand{\ssfr}{sSFR\xspace}
\newcommand{\ssfrs}{sSFRs\xspace}
\newcommand{\sfr}{SFR\xspace}
\newcommand{\sfrs}{SFRs\xspace}
\newcommand{\sfh}{SFH\xspace}
\newcommand{\pz}{photo-$z$\xspace}
\newcommand{\pzs}{photo-$z$'s\xspace}
\newcommand{\eg}{e.g.,\xspace}
\newcommand{\ie}{i.e.,\xspace}

\newcommand{\knn}{$k$-NN\xspace}
\newcommand{\biasz}{\text{bias}_z}

\newcommand{\nsmall}{\num{7799}\xspace}  
\newcommand{\nlarge}{\num{603680}\xspace} 
\newcommand{\ntotal}{\num{611479}\xspace} 

\newcommand{\magfamily}[1]{\texttt{#1}\xspace}
\newcommand{\modelmag}{\magfamily{modelMag}}
\newcommand{\psfmag}{\magfamily{psfMag}}
\newcommand{\devmag}{\magfamily{deVMag}}
\newcommand{\fibermag}{\magfamily{fiberMag}}
\newcommand{\petromag}{\magfamily{petroMag}}
\newcommand{\expmag}{\magfamily{expMag}}

\usepackage[usenames,dvipsnames,svgnames,table]{xcolor}

\newcommand{\revcolor}{}
\newcommand{\rev}[1]{{#1}}



\newcommand{\tssfrrmsea}{\SI{2.90(18)e-01}{log(yr^{-1})}\xspace}

\newcommand{\tssfrrmseb}{\SI{2.71(15)e-01}{log(yr^{-1})}\xspace}

\newcommand{\tssfretab}{\SI{1.72(32)e+00}{\percent}\xspace}
\newcommand{\tssfrrmsec}{\SI{3.49(16)e-01}{log(yr^{-1})}\xspace}

\newcommand{\tssfretac}{\SI{3.05(35)e+00}{\percent}\xspace}
\newcommand{\bssfrrmsea}{\SI{3.04(6)e-01}{log(yr^{-1})}\xspace}

\newcommand{\bssfrrmseb}{\SI{2.83(4)e-01}{log(yr^{-1})}\xspace}

\newcommand{\tzsignmada}{\num{1.72(10)e-02}\xspace}

\newcommand{\tzsignmadb}{\num{1.45(7)e-02}\xspace}

\newcommand{\tzsignmadc}{\num{1.54(9)e-02}\xspace}

\newcommand{\bzsignmada}{\num{1.83(3)e-02}\xspace}

\newcommand{\bzsignmadb}{\num{1.40(2)e-02}\xspace}

\newcommand{\bzsignmadc}{\num{1.65(2)e-02}\xspace}

\title[Selecting bands for optimal accuracy]{Sacrificing information for the
greater good: how to select photometric bands for optimal accuracy}

\author[K. Stensbo-Smidt et al.]{Kristoffer Stensbo-Smidt$^{1}$,
Fabian Gieseke$^{1}$,
Christian Igel$^{1,3}$,
Andrew Zirm$^{2}$,
\newauthor and Kim Steenstrup Pedersen$^{1,3}$\\
$^{1}$Department of Computer Science,
  University of Copenhagen\\
$^2$Dark Cosmology Centre, Niels
  Bohr Institute,  University of Copenhagen\\
$^3$Space Science Center, University of Copenhagen
}

\date{Accepted XXX. Received YYY; in original form ZZZ}

\pubyear{2016}

\begin{document}

\label{firstpage}
\pagerange{\pageref{firstpage}--\pageref{lastpage}}
\maketitle

%

\begin{abstract}
  Large-scale surveys make huge amounts of photometric data
  available. Because of the sheer amount of objects, spectral data
  cannot be obtained for all of them. Therefore it is important to
  devise techniques for reliably estimating physical properties of
  objects from photometric information alone. These estimates are
  needed to automatically identify interesting objects worth a
  follow-up investigation as well as to produce the required data for
  a statistical analysis of the space covered by a survey.  We argue
  that machine learning techniques are suitable to compute these
  estimates accurately and efficiently.  This study
  \rev{promotes a feature selection algorithm, which
  selects the most informative magnitudes and colours for a given task of
  estimating physical quantities from photometric data alone. Using $k$ nearest
  neighbours regression, a well-known non-parametric machine learning method, we
  show that using the found features significantly increases the accuracy of the
  estimations
  compared to using standard features and standard methods. We illustrate the
  usefulness of the approach by estimating specific
  star formation rates (\ssfrs) and redshifts (\pzs) using only the broad-band
  photometry from the 
  Sloan Digital
  Sky Survey (SDSS). For estimating \ssfrs, we demonstrate that our method
  produces better estimates than traditional spectral energy distribution
  (SED) fitting. For estimating \pzs,
  we show that our method produces more accurate \pzs than the method employed
  by SDSS.}
  The study highlights the general
  importance of performing proper model selection to improve the
  results of machine learning systems and how feature selection can
  provide insights into the predictive relevance of particular input
  features.
\end{abstract}

\begin{keywords}
    galaxies: distances and redshifts --
    galaxies: star formation --
    galaxies: statistics --
    methods: data analysis --
    methods: statistical --
    techniques: photometric
%
%
\end{keywords}

%
\section{Introduction}
High-resolution spectroscopic data contain a wealth of information about
astrophysical objects. Analyses relying on spectroscopy suffer, however,
from small sample sizes. Photometric surveys have the
potential to overcome this limitation, but are limited in
terms of the amount of information that can be extracted for each astrophysical
object.
Due to the abundance of data currently available, and especially with
the surveys commencing within the next decade, methods are required
that can automatically extract relevant information from the
broad-band images of these surveys.  Our goal is to reliably,
efficiently and accurately estimate properties of objects from
photometric data, for example, for quickly identifying interesting
objects worth a follow-up investigation or for conducting large-scale
statistical analyses.  In this study, \rev{we apply a method for selecting
the most informative colours and bands for photometric estimations. We
illustrate its potential by estimating specific star-formation rates (\ssfrs)
and photometric redshifts (\pzs) from available SDSS data, but the method can
readily be applied to other quantities and surveys.}

\subsection{Star formation rates}
An ongoing quest in cosmology is the understanding of galaxy formation and
evolution. A crucial part here is to understand the star formation history of
the individual galaxies as well as the universe as a whole. Major open questions
include which processes trigger star formation and, equally important, quench it.
Data from large surveys, such as the Sloan Digital Sky Survey
\citep[SDSS,][]{York2000}, have shown a peculiar bimodality in the star-formation
rates (\sfrs) of galaxies \citep{Kauffmann2003b}. \rev{The bimodality points
to a scenario where star formation is quenched, but the responsible mechanism is
far from understood. Current results indicate that the quenching time-scale
varies significantly with galaxy mass \citep{Wetzel2012,Wetzel2013,Wheeler2014}
and redhift \citep{Balogh2016}, suggesting tha different processes are in play at
different times and masses
\citep{Fillingham2015,Wetzel2015}.
}
%
To uncover these processes, it is natural to turn to the
statistical properties of a large number of galaxies in order to look for
correlations between \sfrs and other physical properties.

The most common way to estimate the recent \sfr of a galaxy is to use a number
of observational tracers.
These tracers often rely on observations of single or multiple emission lines,
with the H$\alpha$ emission line being among the most popular
\citep{Kennicutt2012}.
A main limitation is that they usually require
high-quality spectra.
Other methods to estimate the \sfr include conversion factors to convert from
flux over a given wavelength interval \citep{Kennicutt1998,Kennicutt2012} and
spectral energy distribution (SED) fitting, which relies on a library of
template spectra generated by stellar population synthesis models \citep[for
recent reviews, see][]{Walcher2011,Conroy2013}. In the most basic version of
this method an observed galaxy spectrum is compared to every template spectrum,
the closest match is chosen and the template's physical properties adopted
\citep[\eg][]{Charlot2002,Brinchmann2004}.
%

\rev{
SED fitting is often considered a less precise way to estimate \sfrs than
relying on observational tracers \citep{Walcher2011}. It does, however, allow
us to estimate the \sfr from broad-band photometry, where observational tracers
have more limited use \citep{Maraston2010}.
}

More direct estimations of \sfrs and specific \sfrs (\ssfrs) from
broad-band photometry have also been investigated
\citep[\eg][]{Williams2009,Arnouts2013}, though there are still significant
discrepancies between these estimated quantities and those obtained from more
reliable methods.

\subsection{Photometric redshifts}
\rev{
Spectroscopic surveys provide highly accurate redshifts of galaxies, enabling a
detailed 3D view of galaxy distribution in the universe, but they are both
expensive and time-consuming.
Photometric surveys, on the other hand, can cover a much larger area of the sky
in less time, and can usually go below the spectroscopic flux limit. They
therefore provide a significantly more complete, and thus less biased, sample
of galaxies, which is a notable advantage over spectroscopic surveys.
Photometric surveys, however, struggle with reduced accuracy in the galaxy
positions along the line of sight. Despite this problem, the larger galaxy
sample sizes are useful for numerous cosmological applications, such as
obtaining constraints on
cosmological parameters
\citep[\eg][]{Padmanabhan2007,Carnero2012,Ho2012}.
These applications rely on photometric redshifts (\pzs) calculated from
broad-band photometry.
Naturally, increasing the accuracy
of \pzs is of great
importance.
}

\rev{
    A vast amount of methods have been developed to estimate \pzs
    \citep[see, \eg][for recent
    comparisons]{Hildebrandt2010,Abdalla2011}.
    Broadly speaking, \pz estimation methods can be classified as either
    template-based or empirical methods. Template-based methods use SED fitting
    in the same way as for \sfr estimation: they match the observed colours or
    magnitudes to those of a large library of synthetic template spectra
    \citep[\eg][]{Benitez2000,Bolzonella2000,Ilbert2006,Brammer2008}.
    Empirical methods train algorithms to estimate \pzs from colours or
    magnitudes. The algorithms are calibrated to fit the task at hand using a
training dataset with spectroscopically derived redshifts.}

\rev{
    A wide range of empirical methods have been developed, and most fall
    into the categories of either tuning the colour-$z$ relation or machine
    learning. The machine learning category is highly diverse, with
    techniques such as artificial neural networks \citep{Collister2004},
    self-organising maps \citep{Geach2012}, random forests
    \citep{CarrascoKind2013}, and Gaussian processes \citep{Almosallam2016}
    having been used for \pz estimation. These techniques generally outperform
    template-based methods for \pz estimation, as machine learning methods are
    able to adapt to the highly nonlinear relation between colours and redshift.
    For recent reviews of the performances of various \pz estimation
    methods, see \citet{Dahlen2013} and \citet{Sanchez2014}.
}

\subsection{Increasing the information from photometric measurements}
\label{sub:increasing_information}
\rev{SED fitting is a common method for both \pz and \sfr estimation. Advantages
of this method include the ability to get the full star formation history (\sfh)
(limited by the detail level of the template library) of a galaxy as well as
constraints on its redshift, environment etc. The restrictions lie in the
generation of the template spectra, with computational power and understanding
of stellar evolution being the main limiting factors.}

The main computational limitation is the enormous amount of free parameters that
can be tweaked in the generation of a single spectrum.
%
Because of this, 
and limited physical knowledge about stellar evolution, it is
still a great challenge to generate appropriate template spectra
\citep[\eg][]{Pacifici2015,Smith2015}.
A brute-force way of
calculating templates for a chosen grid of parameters quickly becomes
infeasible.
The amount of degeneracies between the evolutionary states of different single
stellar populations (SSPs) also limits this approach.

A number of ways to reduce the amount of necessary template spectra with minimum
information loss have been explored.
In particular, machine learning methods have been used to interpolate between
template spectra to allow for a sparser grid to be sampled
\citep[\eg][]{Tsalmantza2007}. Active learning was explored by
\citet{Solorio2005}, where the computer automatically generates new
template spectra if no close match is found in the dataset. This automatically
refines the template grid in regions that have actual observations. A different
approach was taken by \citet{Richards2009a}, who used diffusion $K$-means to
tackle the problem of choosing which SSPs make up a
galaxy spectrum, by finding an appropriate basis from a large set of SSP
spectra. In the same spirit, \citet{Chen2009} used
a principal component analysis to estimate specific \ssfrs from obtained
eigenspectra.

\rev{While spectroscopy
    is superior in terms of information content,
photometry excels in terms of coverage.}
Using a machine learning approach to estimate parameters can give us the best of
both worlds. The algorithm can be trained on galaxies with accurate
\rev{parameters} determined from high-resolution spectra and then be used to
estimate the same \rev{parameters} of other galaxies from broad-band photometry
only.  This avoids the problem of generating template spectra from models that
may suffer from various restrictions and approximations.
\rev{However, just as template-based methods
require the parameter space to be densely sampled in order to provide good
parameter estimations, machine learning methods require training data that
represent the entire population. If such are not available, the methods may lead
to biased estimates.}
Machine learning methods can also achieve significantly lower computational
complexity compared to SED fitting\rev{, depending on the level of detail
wanted}, which will become increasingly important in the near future, when new
photometric surveys start producing data at an unprecedented rate.

\rev{Using highly detailed data can, however, lead to a decrease in accuracy.
This counter-intuitive phenomenon occurs for both template methods as well as
machine learning methods,
and can attributed to the fact that
if a dimension contributes only (or even just some) noise, it
will decrease
the overall signal-to-noise ratio (S/N).
}

\rev{Selecting only the most informative dimensions of the data can therefore
lead to higher accuracy, even if it requires removing somewhat informative
dimensions, as the lower dimensionality of the data
can result in a higher S/N.
}

\rev{In the machine learning literature, the dimensions of a data point are
referred to as \emph{features}. Thus, the task of choosing the most informative
dimensions is called \emph{feature selection}. Feature selection has already been
investigated in an astrophysical context. Among the most used feature selection
algorithms are random forests, which produce feature ranking as part of the
algorithm. They have been used in a number of studies, for example,
\citet{DIsanto2016} and \citet{Rimoldini2012}.
Random forests are not the only way to select features, and
\citet{Graham2013}
tested five different feature selection strategies for classifying stars.
\citet{Hoyle2015} showed how adding the most informative features to the
standard set of colours and magnitudes significantly increased the accuracy for
\pz estimation.}

\rev{It is important to realise that the concept of most informative features is
not a universal one; the most informative features for one algorithm may be
different from those of another. That depends on how specifically the algorithm
uses the features, for example, some algorithms may be sensitive to scaling of the
features, while others may not. And just as the most informative features vary
from algorithm to algorithm, so will they vary from task to task. For example,
whereas observed UV radiation may contain a lot of information regarding star
formation in the nearby universe, it may not be that informative for
detecting, say, brown dwarfs.
}


In this paper, we show that we can obtain a significantly greater accuracy of
\rev{estimated \pzs and \ssfrs}, using only SDSS \emph{ugriz} photometry, by
applying a machine learning method rather than relying on spectral modelling of
the photometry.
\rev{Our approach is similar to that of \citet{Stensbo-Smidt2013}, but here we
show that the accuracy can be further increased by performing a feature
selection, selecting the most informative features among all measured SDSS
magnitudes and colours.}

Specifically, we use $k$-nearest neighbours (\knn) regression, which
is an intuitive method well-known in machine learning and to some extent also in
astronomical communities \citep[see,
\eg][]{Li2008,Polsterer2013,Polsterer2014,Kugler2015,Kremer2015}.
\rev{Of the more prominent uses of \knn in astronomy is the estimation of \pzs
in SDSS \citep{Abazajian2009}.}

By using \knn we can automatically learn a mapping from magnitudes and colours
of galaxies to their parameters derived from reliable indicators, thereby
allowing accurate photometric estimates
without high-resolution spectra.
The reliable parameters can be estimated using any method deemed appropriate for
each individual galaxy, effectively taking advantage of multiple indicators, as
explored by \citet{Wuyts2011,Wuyts2013} for \sfrs. A significant advantage of
\knn over other methods is that it naturally adapts to the local, potentially
high-dimensional structure of the data, and can thus model highly non-linear
behaviour without problems.
\rev{Another virtue of \knn is its simplicity, which makes it easy to see how
    data are used and compared within the algorithm.
}


\rev{Selecting the most informative features can, in theory, be done by trying
all possible feature combinations. As the number of combinations grows
exponentially with the number of features, this quickly becomes unfeasible, and
one has to resort to clever selection strategies. Here, we use \emph{forward
feature selection} to determine the most informative features
(see section~\ref{sub:feature_selection} for details). Forward feature selection
was used by \citet{Xu2013} to examine which halo properties contained most
information about the number of galaxies. In this paper, we use it to improve
the estimation of \pzs and photometric \ssfrs, which illustrate the
method's general usefulness.}


%


The remainder of this paper is organised as follows: in section~\ref{sec:method}
we describe the \knn algorithm and the algorithm we use to select the most
informative colours. Section~\ref{sec:experimental_setup} describes the data we
are using and details our experimental set-up. In section~\ref{sec:results} we
provide results of our experiments and an analysis of these.  We end with a
discussion and a summary of our conclusions in
section~\ref{sec:conclusion}.

\newcommand{\Reals}{{\mathbb R}}
\newcommand{\tsize}{N}
\newcommand{\numfeat}{\bar{d}}
\newcommand{\tdim}{D}
\newcommand{\kdtree}{\mbox{\ensuremath{k}-d}~tree\xspace}
\newcommand{\kdTree}{\mbox{\ensuremath{k}-d}~Tree\xspace}
\newcommand{\kdTrees}{\mbox{\ensuremath{k}-d}~Trees\xspace}
\newcommand{\kdtrees}{\mbox{\ensuremath{k}-d}~trees\xspace}
\newcommand{\Kdtree}{\mbox{\ensuremath{K}-d}~tree\xspace}
\newcommand{\KdTree}{\mbox{\ensuremath{K}-d}~Tree\xspace}
\newcommand{\KdTrees}{\mbox{\ensuremath{K}-d}~Trees\xspace}
\newcommand{\Kdtrees}{\mbox{\ensuremath{K}-d}~trees\xspace}
\newcommand{\GPU}{{\sc GPU}}
\newcommand{\CPU}{{\sc CPU}}
\newcommand{\GPGPU}{{\sc GPGPU}}
\newcommand{\GPUs}{{\sc GPUs}}

\section{Methods}
\label{sec:method}

The goal of this study is to \rev{test the efficiency of machine learning
    techniques, in particular
    feature selection,
    when estimating physical quantities of galaxies. We
    suggest using the
selected features directly in regression methods rather than in
connection with physical models, such as population synthesis models.}
There are two fundamental ways of doing regression:
parametric and non-parametric. In the parametric case, data is assumed to
follow a function $f(x)$ with known form but unknown parameters. It is usually
fairly easy to estimate these parameters by fitting, but this advantage comes at
a cost: by choosing a particular functional form of $f(x)$ we have made
assumptions about the underlying structure of the data. If these assumptions are
not absolutely correct, we will not be able to achieve optimal estimation
performance \citep{James2013}.  This is where non-parametric methods have an
advantage, as they do not make any assumptions about the structure of the data,
but adapt to it.

\subsection{$k$ nearest neighbours regression}
We employ one of the simplest non-parametric methods, namely
    \emph{$k$ nearest neighbours (\knn) regression}
\citep{Altman1992,Hastie2009,James2013}. Assume that we are given a data set
$\mathcal S = \{(\vec x_1, y_1), \ldots, (\vec x_\tsize, y_\tsize) \}
\subset \Reals^\tdim \times \Reals$ consisting of $\tdim$-dimensional
data points $\vec{x}_i$ with associated output values $y_i$. For
instance, each data point could represent a galaxy with $\tdim=2$
colour values (\eg $B-V$ and $U-B$) and the output value $y_i$
could be the \ssfr that one is interested in estimating.  The
components of $\vec x_i$ (which, in this example, would be the
colours) are called \emph{features}.
Now we employ machine learning to infer from $S$ a general rule of
how to predict the (unknown) output value~$y'$ given
some new data point $\vec x'$.
The \knn method does this by
simply finding the $k$ closest data points with known output values, and then
taking the average of these values, \ie
%
\begin{align} \label{eq:knn} y' = \frac{1}{k} \sum_{i\in\mathcal N_k}
    y_i\,, \end{align}
where $\mathcal N_k$ is the set of the $k$ nearest data points in
$\{\vec{x}_1,\ldots,\vec{x}_N\}$ w.r.t.\ the new sample $\vec x'$.
The `closeness' between samples is
defined via a metric $d$.
That is, $\mathcal N_k = \mathcal N_{k-1} \cup
    \operatorname{argmin}_{(x,y)\in S\setminus \mathcal
        N_{k-1}}d(\vec{x},\vec{x'})$ for positive integers $k$ and $\mathcal
    N_0=\emptyset$, where $\operatorname{argmin}$ breaks ties at random.
We use the Euclidean metric
$d(\vec{x},\vec{z})=\sqrt{\sum_{i=1}^{\tdim} {(x_i-z_i)}^2 }$ for
$\vec{x},\vec{z}\in\Reals^\tdim$\rev{, though any metric can be chosen. The
    Euclidean distance is the most common choice in the literature, but it is
    perfectly possible that another metric would perform better. One can also
attempt to learn the metric from the data as done by, \eg
\citet{Weinberger2009}. To keep things simple, however, we stick to the
Euclidean metric.}

Although the \knn regression method is simple, it often yields highly accurate
predictors. This is especially the case if the amount of training data $N$ is
large and/or the feature space dimensionality $D$ is low.
\rev{While it may seem counterintuitive, adding more features (\ie
dimensions) to the input data may make \knn perform worse.}
The performance of
nearest neighbours models can deteriorate if $D$ gets too large,
\rev{in particular when each added dimension contains intrinsic noise. The
addition of extra noise with each added dimension may eventually decrease the
S/N.
This is perhaps most easily recognised if one considers the extreme case of
adding a feature, which is pure noise. This can only decrease the performance,
and adding more of these pure noise features will eventually down any signal
present in the original features.
}
Thus, it is important to select the right features, see
section~\ref{sub:feature_selection}.



\revcolor
\subsubsection{Dealing with uncertainties}
\label{sub:uncertainties}

In its most basic form, the \knn algorithm does not support the inclusion of
uncertainties associated with inputs or outputs, nor does it provide
confidence intervals for the estimated quantities beyond calculating the
variance of the neighbours' outputs \citep{Altman1992}.
There are, however,
extensions dealing with these issues.

There are a number of ways uncertainties
may influence the results of an analysis.
Firstly, there may be uncertainties related to the output values (\eg \ssfrs or
\pzs) of the training data, which need to be propagated
to the
predicted output. Secondly, there may be uncertainties in the
input values (\eg colours) of both the training data and the new data, which also
need to be propagated to
the estimated output value.

Propagating uncertainties from known data to the estimate done by \knn is not a
trivial task.
Ideally, to estimate the output value of a new datum, its input uncertainties
need to be propagated, and one needs to incorporate the uncertainties on both
input and output of the training data.
%
A standard Monte Carlo sampling can deal with all these uncertainty issues, but
it will quickly get far too computationally expensive.


Assuming Gaussian errors, uncertainty in the output alone can be dealt with in
a relatively straightforward manner by using a weighted average, $y' =
\sum_{i\in\mathcal N_k} w_i y_i/\sum_{i\in\mathcal N_k} w_i$, using $w_i
= \sigma_i^{-2}$, where $\sigma_i^2$ is the variance of $y_i$. This does not
account for the
scatter of the inputs, which ideally should mean that more distant neighbours
(and their corresponding uncertainties) are weighted less when computing the
average. This can be accounted for by including the similarity metric in the
weights, or, alternatively, including the uncertainties in the similarity
metric as done by, \eg \citet{Polsterer2013}.
An additional complication arises due to the uncertainties in
the inputs and the choice of number of neighbours, $k$. With uncertain inputs,
the question of which of two neighbours is closer cannot be answered with
complete certainty.

Both the question regarding choosing $k$ and that of choosing the proper
similarity metric can, however, be addressed with a probabilistic formulation of
\knn
%
\citep{Holmes2002,Everson2004,Manocha2007}, which allows for posterior inference
over $k$ and the similarity metric.


Finally, one may simply try to find a heuristic, reasonable estimate of the
uncertainty of the new data. This is for instance how the \pz uncertainties in
the SDSS database have been
computed \citep{Abazajian2009}.
Here, a hyperplane was fitted to the nearest 100 neighbours in colour space, and
the mean deviations of the redshifts from this hyperplane were found to be good
estimates of the errors.

To our knowledge, there is no accepted way of dealing with all uncertainty issues
short of Monte Carlo sampling. In this paper, we have therefore chosen to ignore
the question relating to uncertainties, focusing solely on demonstrating the
performance gain of combining \knn and feature selection.

\color{black}

\subsection{Choosing the number of neighbours}
\label{sub:choosing_the_number_of_neighbours}

\rev{In most versions of \knn, including the vanilla version, one much choose
the number of neighbours, $k$, to average over.}
Increasing $k$ implies that a prediction will be based on the average of many
samples, which reduces the variance of the classifier but may increase
its bias \citep[for a discussion of the bias-variance decomposition of the
error of \knn regression we refer to][]{Hastie2009}.
A standard technique for choosing $k$ is
cross-validation (CV). In $M$-fold CV, the available data $S$ is
randomly partitioned
into $M$ subsets $S_1,\dots, S_M$ of (almost) equal size.
Let $S_{\setminus i}=\bigcup_{j=1,\dots,M \wedge j\neq i} S_j$
denote all data points except those in $S_i$.
For each $i=1,\dots,M$, an individual model is built by applying the algorithm
to the training data $S_{\setminus i}$. This model is then evaluated using the
test data in $S_i$.  The average error is called \emph{cross-validation
error} and is a predictor of the generalisation performance of the algorithm.
To choose $k$ for \knn using $M$-fold CV, $S$ is split into $M$ subsets. For
each fold $i=1,\dots,M$ \knn models are built and tested using different values
for $k$ (say, $k=1,3,5,\dots$). The $k$ with the lowest CV error is finally
selected.

It must be stressed that the data used for model selection must be independent
from data for assessing the final performance of a model.

\subsection{Informative features}
\label{sub:feature_selection}

The use of appropriate features is crucial for machine learning.
Standard features in astronomy are, for instance, magnitudes or the derived
colours. \rev{The performance of a model can, however,} often
be improved by considering additional features or special combinations of
features (thus, effectively changing the underlying distance
metric $d$).\footnote{For instance, the SDSS pipeline resorts to two different types
    of magnitudes via the linear model $\texttt{psfMag}-\texttt{cModelMag} >
    0.145$ to classify photometric objects as `galaxy' or
    `point-like'.}
We employ automatic \emph{feature selection} to pick the most
informative features for our regression task.

\subsubsection{Feature selection}
The goal of feature selection is to reduce the
dimensionality of the input space by selecting the most informative
features.
A direct way to select such informative features is to systematically
try various combinations of features and select the subset with the
most promising accuracy for the final model (based on a certain evaluation
criterion such as CV). In theory, one would like to try every possible
combination of features, but in practice this is often infeasible due to the
induced exponential runtime. In the literature, different techniques have been
proposed to address this issue such as the idea to maximise the probability of
finding the best combination of features. We refer to \cite{Guyon2003} for an
introduction to feature selection.

Standard alternatives to such an exhaustive search are \emph{forward} and
\emph{backward feature selection}~\citep{Hastie2009}, which aim at selecting
informative features in an incremental manner. For the case of forward
selection, one starts by selecting the most promising feature by assessing the
predictive power of each of the $\tdim$ features. In the second iteration, the
first feature is kept and a second one is selected based on the predictive power
of both the first \emph{and} the second feature. This process is repeated until
the number $\numfeat$ of desired features is selected. Backward elimination
works similarly. However, instead of incrementally adding features, one removes
a feature at a time, starting with all $\tdim$ features being selected.

\rev{Even forward and backward feature selection are still computationally
demanding, but using clever implementations and data structures one may
parallelize the procedure. This paper uses a massively-parallel matrix-based
implementation combining incremental feature selection and nearest neighbour
models, recently proposed by \cite{GiesekePOI2014}. For more details on the
implementation, we refer to appendix~\ref{app:feature_selection}.}

\section{Experimental set-up}
\label{sec:experimental_setup}

\subsection{Data selection}
\label{sub:data_selection}

\begin{figure}
    \centering
    \includegraphics[width=0.9\columnwidth]{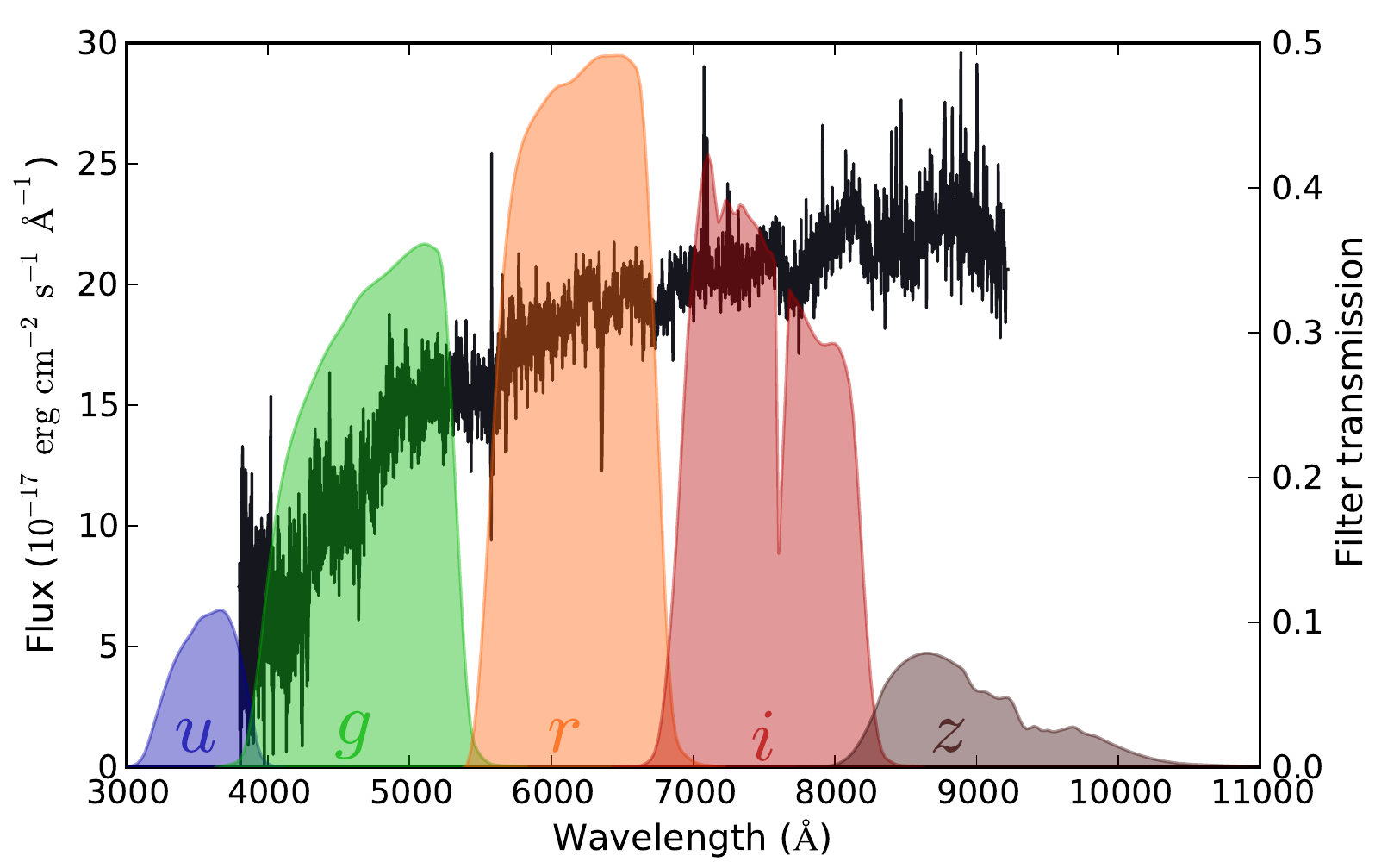}
    \caption{An example spectrum of a galaxy from the SDSS database (black
        curve) overlaid by the five bandpass filters of SDSS \citep{Fukugita1996}.}
    \label{fig:sdss_bands}
\end{figure}

The experiments in this study use photometric data from Sloan Digital Sky
    Survey \citep[SDSS,][]{York2000}. The data are a subset of SDSS Data Release 7
\citep[DR7,][]{Abazajian2009}, and consist of \psfmag,
\fibermag, \petromag, \devmag, \expmag, and
\modelmag magnitudes in the \emph{u, g, r, i,} and \emph{z} bands (see
Fig.~\ref{fig:sdss_bands}) for
each galaxy as well as
the galaxy's \ssfr\ \rev{and redshift}, estimated from
spectroscopy.
\rev{We also include the photometric redshifts estimated by SDSS
\citep{Abazajian2009}.}

\rev{Data are obtained from SDSS CasJobs, using the \texttt{SpecPhoto}
view, which ensures that objects have clean spectra. Specific star formation
rates were taken from \citet{Brinchmann2004}\footnote{We used
\texttt{specsfr\_avg} from the data located at
\url{http://wwwmpa.mpa-garching.mpg.de/SDSS/DR7/sfrs.html}}.
To clean the data, we apply the following constraints:
\begin{itemize}
    \item For \ssfrs, we require that the estimation was successful
        (\texttt{flag = 0}), and we remove all duplicate galaxies.
    \item For redshifts, we require that both spectroscopic and photometric
        estimations were successful (for spectroscopy, \texttt{zWarning = 0}; for
        photometry, \texttt{zErr >= 0}).
\end{itemize}
}

A sample of \ntotal galaxies meet the above criteria. For a smaller subset of
\nsmall low-redshift galaxies ($0.0042<z<0.33$) within the selected sample, we
additionally have \rev{photometric \ssfr estimations} obtained by a
template-based modelling approach described in
section~\ref{sub:estimations_from_galaxy_spectra_templates}. No additional
selection criteria have been applied to this subset.
\rev{In particular, no S/N cut has been used in order to
    highlight the method's robustness to varying noise levels. In this work, we
    do not make use of any S/N information, though special treatment of low S/N
    sources can be incorporated in various ways (see discussion in
section~\ref{sub:uncertainties}).}

\begin{figure*}
    \centering
    \subfloat[Small subset.\label{fig:z:template}]
        {\includegraphics[width=0.48\textwidth]{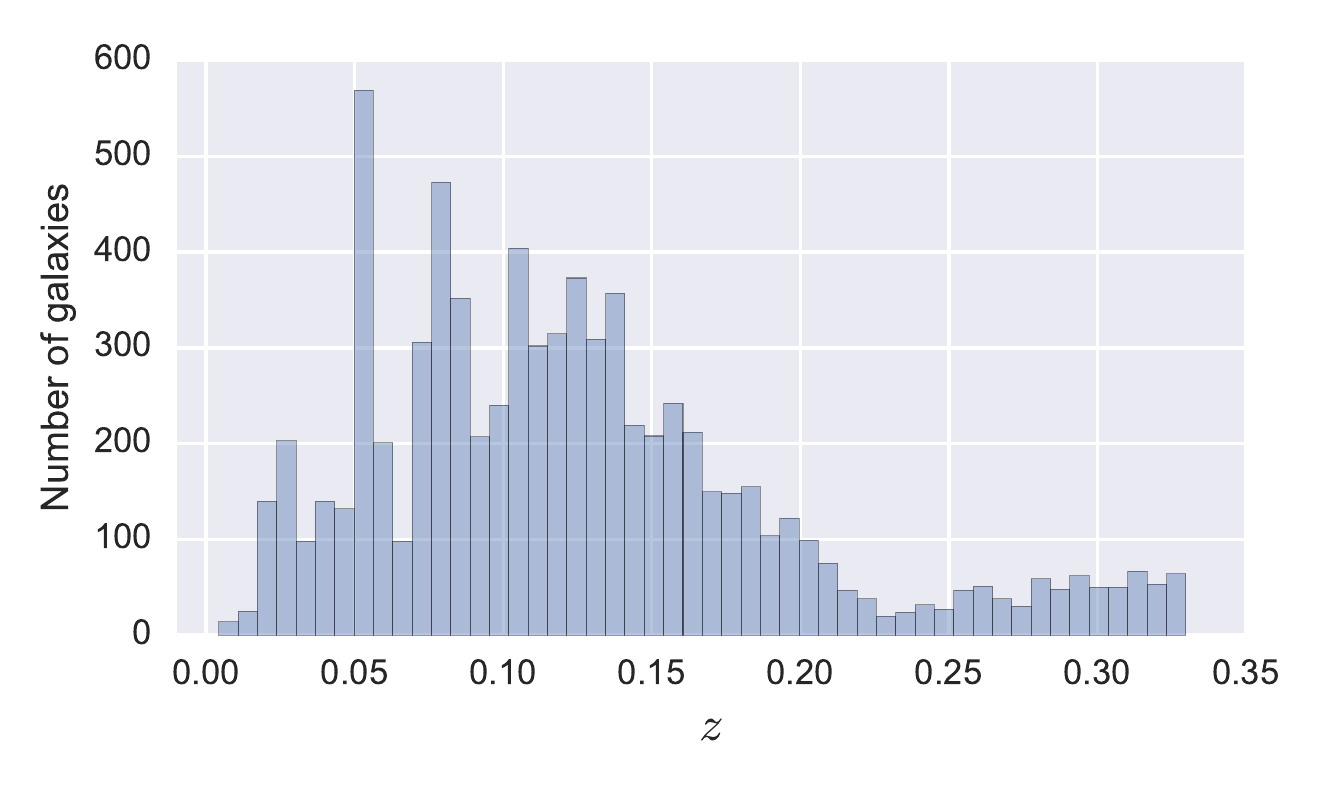}}
        \hfill
    \subfloat[Entire sample, exluding smaller subset.\label{fig:z:bigdata}]
        {\includegraphics[width=0.48\textwidth]{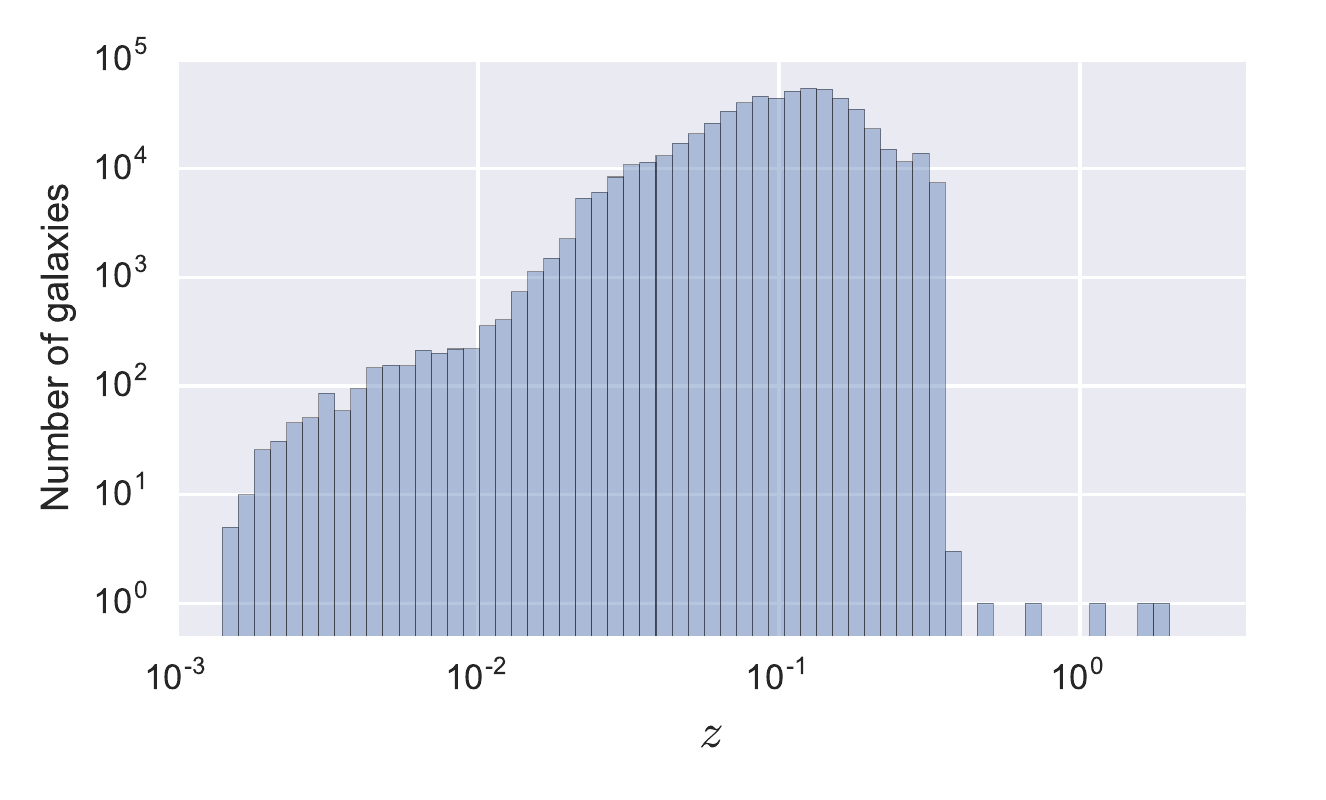}}

    \caption{Redshift distribution of the two galaxy samples used in the
    experiments. The entire sample, excluding the smaller subset, additionally
    contains a single galaxy with $z = -\num{1.93e-4}$, which is not shown in
the plot.}
    \label{fig:redshift_distribution}
\end{figure*}
The experiments will be based on two samples of galaxies: the smaller subset
and the full sample, excluding the smaller subset (totalling \nlarge galaxies).
The redshift distributions of these two samples can be seen in
Fig.~\ref{fig:redshift_distribution}. The smaller subset consists entirely of
low-redshift galaxies, where also the majority of the larger sample can be
found. The larger sample \rev{consists primarily of galaxies below $z \sim 0.5$,
with only a few galaxies at higher redshifts.}

\subsection{\rev{Comparison with other methods}}
\label{sub:estimations_from_galaxy_spectra_templates}

\rev{We will compare our results to those of two other methods, one for the
\ssfr estimations and one for the \pz estimations.

For the \pz experiments, we
compare our results to the \pzs available directly through the SDSS database.
These \pzs have been estimated using a combination of \knn \citep{Csabai2007}
and a template-based method \citep{Budavari2000}, as described in
\citet{Abazajian2009}. The \knn part of the method differs from our approach in
that it bases the estimated \pz on
a local hyperplane fitted to the 100 nearest
neighbours, instead of just taking the average (and optimizing the number of
neighbours), as we do. It is also important to
note that our experimental set-up is different to the one SDSS uses. In
particular, SDSS have likely based their \pz estimates on a much larger
training set than we use.

For the \ssfr experiments,}
we compare our estimated \ssfrs to those obtained by the standard approach of
stellar population synthesis modelling very similar to those used in
\citet{Gallazzi2005,Gallazzi2008,Salim2007}\footnote{J. Brinchmann, private
communication.}.
%
Roughly speaking, a large library of template spectra is generated from stellar
population synthesis models.
To estimate the \sfr of a certain observed galaxy, one would compare the galaxy's
spectrum to each of the template spectra.  The \sfrs of the templates are then
weighted based on the likelihood of the template spectra given the real
spectrum, resulting in a probability distribution for the \sfr.  From this
distribution, the final \sfr of the galaxy is calculated as the expected \sfr.

\rev{To estimate the \ssfr when only photometric information is available, the
template spectra are multiplied by the filter transmissions of the particular
survey, in our case SDSS (see Fig.~\ref{fig:sdss_bands}), to produce template
magnitudes. These are then compared to observed ones, and the pipeline described
above continues.}




\subsection{\rev{Description of experiments}}
\label{sub:experiments}

We considered four different experimental set-ups with the common goal of
estimating \ssfrs\ \rev{and \pzs} of galaxies as accurately as possible. The first two
experiments were based on the exact same galaxy sample as used for the
template-based model (and can thus be compared directly), whereas the last two
experiments were based on the total selected galaxy sample mentioned in
section~\ref{sub:data_selection}, but with the smaller subset excluded
(hereafter referred to as the \emph{larger subset}). \rev{The experiments for
\ssfr and \pz estimations were identical in set-up -- only the quantity to be
estimated changed.}

Common to all experiments is that we used the four colours $u-g$, $g-r$, $r-i$,
and $i-z$ of the galaxies, \rev{and in the experiments with feature selection we
additionally included the plain magnitudes $u$, $g$, $r$, $i$, and $z$}.
\rev{The magnitudes
varied from experiment to experiment,
see the summary in Table~\ref{tab:experiments} and detailed description further
down. The data and code used for the experiments can be found online, see
appendix~\ref{sec:code_and_data}.}

\begin{table*}
    \newcommand{\hs}{\hspace{3mm}}
    \begin{tabular}{c  c  c  c}
        \toprule
        Experiment   & {Sample size}    & Feature selection & Features \\
        \midrule
            1    &   \nsmall            &   No  & \modelmag; colours only \\
            2    &   \nsmall            &   Yes & \psfmag, \fibermag,
            \petromag, \devmag, \expmag, \modelmag; colours and magnitudes \\
            3    &   \nlarge            &   No  & \modelmag; colours only \\
            4    &   \nlarge            &   Yes & As selected in experiment
            2\\

        \bottomrule
    \end{tabular}
    \centering
    \caption{Summary of experiments. The experiments were based on the four
        colours $u-g$, $g-r$, $r-i$, and $i-z$, \rev{as well as the five magnitudes $u$,
    $g$, $r$, $i$, and $z$, where indicated.}}
    \label{tab:experiments}
\end{table*}

In each experiment, a nested cross-validation (CV) -- an inner and an outer --
was used to asses the performance of the \knn method.
Both the inner and outer
CV partitioned the data
into 10 folds with 9 folds being used for training and the remaining fold being
used for testing.
For each outer CV,
the 9 folds of training data were further partitioned into 10 inner folds
for the inner CV.
Of these 10 inner folds, 9 were used as training data and the remaining as
test data in order to
determine the optimal $k\in\{2,3,4,\ldots,50\}$,
while simultaneously doing feature selection \rev{by minimizing the root-mean-square
error (RMSE).
The exact number of chosen features, as well as which features were chosen,
therefore varied across all folds.}
This simultaneous $k$ determination
and feature selection was made possible by the massively parallel GPU
implementation of the \knn algorithm described in \citet{GiesekePOI2014}.
Doing feature selection on the scale of this
study is simply not feasible without a highly optimized \knn implementation.

After the optimal features and optimal $k$ were determined by the inner CV, the
performance was assessed by the outer CV.

\rev{The performance of each method was therefore assessed 10 times, allowing us
to calculate both the means and (population) standard deviations for each of the
performance metrics discussed in section~\ref{sec:results}.}
As folds in a CV procedure are not fully independent of each other, these
standard deviations cannot be interpreted as strict confidence intervals.

To make the estimations by the \knn, the template-based model \rev{(for the
\ssfr estimations) and the SDSS method (for the \pz estimations)} comparable, the
predictions by the \rev{latter two methods} were
divided into the same 10 subsets as used in the outer CV of the \knn, and the
same statistics were calculated. The four experiments were devised as follows:

\paragraph*{Experiment 1}
\label{par:experiment_1}

The first experiment used the smaller subset (\nsmall galaxies) and used the four
\modelmag colours $u-g$, $g-r$, $r-i$, and $i-z$ as features. No feature
selection was performed, but $k$ was still optimized in each of the inner CV
folds. This experiment acts as a baseline for the later feature selection.

\paragraph*{Experiment 2}
\label{par:experiment_2}

The second experiment again used the smaller subset, but this time all six types
of magnitudes (\psfmag, \fibermag, \petromag, \devmag, \expmag, and \modelmag)
were used. Each type of magnitude gives rise to four colours \rev{and five
    magnitudes, totalling 54 features}. A feature selection was performed
    \rev{independently for each outer CV fold} to find the
best feature combination.

\paragraph*{Experiment 3}
\label{par:experiment_3}

The third experiment used the larger subset. The features were again only the
four \modelmag colours, and the experiment will serve as a baseline for the \knn
performance on this larger subset.

\paragraph*{Experiment 4}
\label{par:experiment_4}

The fourth experiment also used the larger subset. The features were chosen to
be the overall most informative ones found in experiment 2, \rev{based on a
median ranking of the importance of each feature across the CV folds}. This last
experiment will test how well \knn, with features found from a feature selection
on a small data set, can extended to a much larger data set, thus assessing its
performance in a `big data' setting.

\newcommand{\dssfr}{\Delta\text{sSFR}}
\newcommand{\ssfrdiff}{\log_{10}(\text{sSFR}_\text{est}/\text{sSFR}_\text{spec})}
\section{Results and analysis}
\label{sec:results}

\subsection{Specific star formation rate experiments}
\rev{We evaluate the \ssfr experiments using the following performance metrics.
In general, we use the logarithm of the ratio of the estimated \ssfr to the
spectroscopically confirmed, $\dssfr \equiv \ssfrdiff$. For each CV fold $m$, we
compute the root-mean-square error (RMSE) as
\begin{align*}
    \text{RMSE} = \sqrt{\frac{1}{|S_m|}\sum_{n \in S_m} \dssfr_n^2}\ ,
\end{align*}
where $S_m$ is the test set.
We also compute the median of $\dssfr$, as well as
the scatter, $\sigma$, defined to be the standard deviation of $\dssfr$ over the
test set.
Lastly, we report the fraction of catastrophic outliers, $\eta$, defined to be
galaxies with $|\dssfr| > 3\sigma$.

Results of the \ssfr experiments can be seen in Table~\ref{tab:ssfr_results}.
The reported values are the means and standard deviations of each performance
metric over the ten CV folds.}

\begin{table*}
    \centering
    \caption{Root mean square errors (RMSEs), medians and scatter of $\dssfr$,
        shown as their mean and standard deviations over the ten CV folds for the
    \knn regressions and the template-based model.}
    \label{tab:ssfr_results}
    \sisetup{
        table-alignment = center,
        separate-uncertainty = true,
        fixed-exponent = -2,
        table-omit-exponent = true,
        table-format = +2.2(3),
	}
    \newcommand{\hs}{\hspace{5mm}}
    \begin{threeparttable}
        \begin{tabular}{l @{\hs} c @{\hs} S[table-format=2.1(2)] @{\hs} S @{\hs}
                S[table-format=2.1(2)]
            @{\hs} S[fixed-exponent=0,table-format=1.2(2)]}
            \toprule
            Experiment   & $D$ & {RMSE/\SI{e-2}{log(yr^{-1})}} &
            {Median/\SI{e-2}{log(yr^{-1})}} &
            {Scatter, $\sigma$/\SI{e-2}{log(yr^{-1})}} &
            {$\eta / \si{\percent}$} \\
            \midrule

            \multicolumn{6}{c}{SDSS subset of \nsmall galaxies}\\
\midrule
1	&	4	&	2.90(18)e-01 & 1.63(120)e-02 & 2.89(17)e-01 & 1.78(36)e+00\\
2	&	8\tnote{a}	&	2.71(15)e-01 & 1.52(99)e-02 & 2.70(14)e-01 & 1.72(32)e+00\\
Template-based model	&		&	3.49(16)e-01 & -1.24(8)e-01 & 3.04(16)e-01 & 3.05(35)e+00\\
\midrule
\multicolumn{6}{c}{SDSS subset of \nlarge galaxies}\\
\midrule
3	&	4	&	3.04(6)e-01 & 1.50(16)e-02 & 3.04(6)e-01 & 1.74(5)e+00\\
4	&	8	&	2.83(4)e-01 & 1.16(8)e-02 & 2.83(4)e-01 & 1.79(4)e+00\\

            \bottomrule
        \end{tabular}
        \footnotesize
        \begin{tablenotes}
            \item[a] Number of features is the median of the ten CV folds.
        \end{tablenotes}
    \end{threeparttable}
\end{table*}

Comparing first the results of the experiments on the smaller subset of SDSS
(experiment~1 and 2) to the result of the template-based model, we see a clear
overall improvement for both experiments. \rev{In particular, the median is much
improved, showing that \knn achieves a lower bias.}

In addition, doing feature selection (experiment~2) rather than simply using the
four \modelmag colours (experiment~1) further improved the estimations, though not as
significant as the differences to the template-based model.

\begin{figure}
    \centering
    \includegraphics[width=\columnwidth]{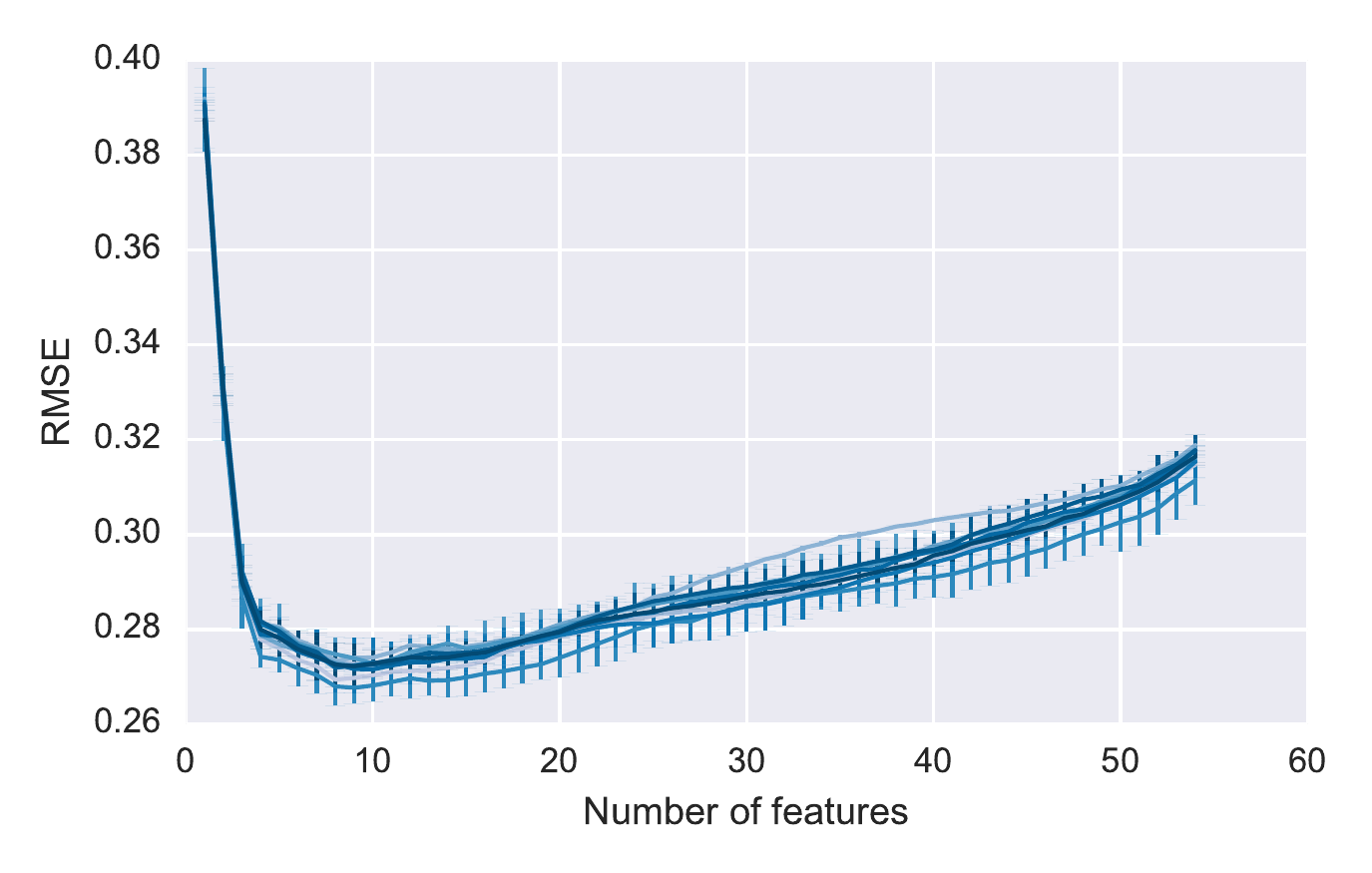}
    \caption{Root mean square error (RMSE) and one standard deviation intervals
        for each of the ten CV folds of the estimated \ssfrs during feature
        selection. A sharp decrease in error is seen as the first features are
        added, but the it levels off quickly after the first three added
        features. As features continued to be added the errors started
        increasing again.
    }
    \label{fig:feature_selection_errors}
\end{figure}
Figure~\ref{fig:feature_selection_errors} shows the RMSE and standard deviation
of the \ssfr estimation \rev{for each of the ten CV folds of the smaller subset
during} feature
selection (experiment~2). The RMSE and standard deviation are computed each time
a feature is added.  It is seen that by far the largest gain in accuracy
happened with the addition of the first three features \rev{(which for all folds
are three \modelmag colours, see below)}.
The error kept decreasing until it was at its lowest at seven to nine added
features after which the error started to increase. This is a very
commonly seen behaviour for \knn, and the reason is likely
the \rev{decreasing} quality of the
features; as the dimensionality of the feature space increases,
we are adding less
informative (\ie noisier) features. The combined effect is that the nearest
neighbours to any given data point might change and the estimation will be worse
as a result.  It is therefore important to stop the feature selection process
before the error starts increasing. The results for experiment~2 in
Table~\ref{tab:ssfr_results} were achieved in exactly this way, \ie by stopping
when the RMSE was lowest.

\begin{figure*}
    \centering
    \includegraphics[width=.9\textwidth]{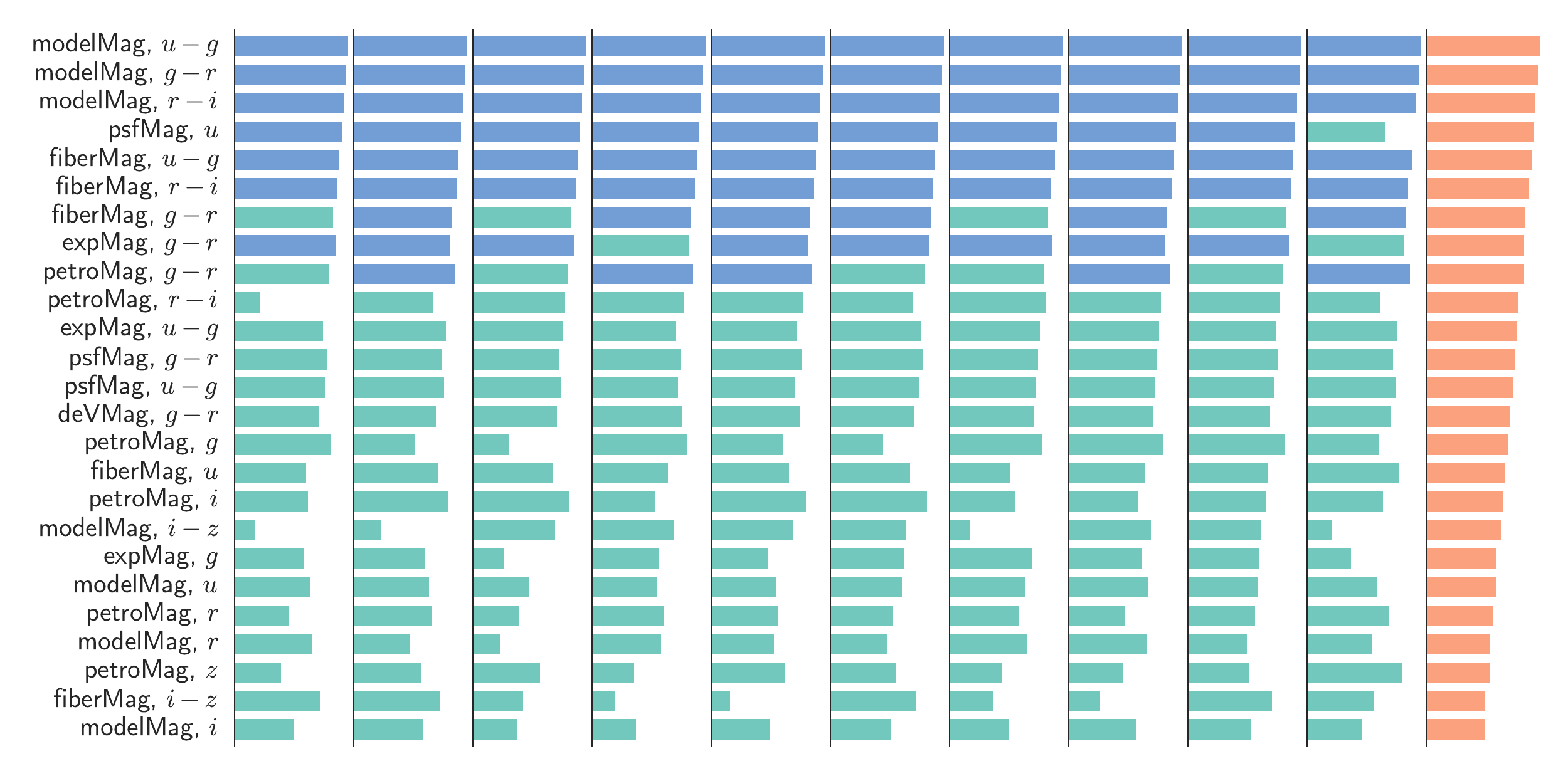}
    \caption{Ranking of the top 25 most important features from the feature
        selection in experiment~2.  To the left are the feature names, while the
        rightmost column shows the median rank of each feature across all CV
        folds. Each of the other columns shows the feature ranking in a
        particular CV fold.  The larger the bar for a certain feature, the more
        important the feature was. Blue bars show features that were chosen
        during the feature selection as the most informative in a particular CV
        fold. Because of the differences in the data used in each CV fold the
        exact features selected as important, as well as the number of chosen
        features per fold, will vary.  The number of chosen features vary
        from 7 through 9, with a median of 8.
    }
    \label{fig:feature_ranking_ssfr}
\end{figure*}

To see which features were chosen in experiment~2, and in which order they were
chosen, the results for each CV are illustrated in
Fig.~\ref{fig:feature_ranking_ssfr}.  The full list of ranked features can
be seen in Fig.~\ref{fig:full_feature_ranking_ssfr}.

The names of the features are shown to the left of the plot, and
the middle ten columns show the ranking of the features for each of the ten CV
folds, with a larger bar indicating \rev{that the feature was chosen
earlier in the feature selection process, and thus has} higher importance. A bar
is coloured blue if the corresponding feature was selected in the feature
selection process. Note that the amount of selected features per fold varies, as
do the chosen features themselves.  This is due to the differences in the data for each
fold. This variation should become less prominent with an increased amount of
data, as the folds will statistically become more and more similar. The
rightmost column shows the median rank of each feature over all CV folds.

It is seen that the top \rev{six} features were consistently chosen in each CV
fold, \rev{except for the $u$ band \psfmag, which was replaced by the $r$ band
in the last fold, see Fig.~\ref{fig:full_feature_ranking_ssfr}}. The remaining
chosen features varied more, but were also consistent enough that, except for
the $r$ band \psfmag, no feature below the \rev{ninth} in the figure was ever
chosen.
\rev{For the overall most informative features (for use in experiment~2 and 4),
we chose to select the top eight features from the plot, since eight is the
median of the number of chosen features across the CV folds. This is
just one particular
choice, and one may
equally well use
other selection criteria or ranking
methods, \eg ranking based on how often a feature was chosen.
Indeed, testing various ranking and selection criteria is an
obvious extension to our work, though the exact choices are unlikely to cause
significant changes to the results.
In summary, we chose to base both
ranking and selection on the median.}

\rev{Returning to the figure, it is interesting that only a single \petromag
colour was ever selected, even though these} are the magnitudes recommended by
the SDSS\footnote{\url{http://classic.sdss.org/dr7/algorithms/photometry.html}}
for use with low-redshift galaxies.
\rev{Instead, the most prominent features were \modelmag and \fibermag colours,
with \modelmag colours as} the top three most informative features. This is not
surprising, since the \modelmag magnitudes are defined as either \expmag or
\devmag magnitudes depending on which fits the best.

\rev{Interestingly, none of the selected features use the $z$ band}, which
can likely be explained by the \rev{band's} low filter transmission, as seen in
Fig.~\ref{fig:sdss_bands}. This will often result in a low signal-to-noise ratio
(S/N).
\rev{Also interesting is the fact that the $u$ band appears in many of the
    most informative features, even though it also has a low S/N. The reason is
likely that the band captures UV radiation from newly formed stars, thus
directly measuring (part) of the \sfr.}

\rev{Looking further down the list of selected features, we see that magnitudes
    and colours based on \expmag were generally ranked much higher than their
\devmag counterparts. This is interesting, as \modelmag, which dominates the
list of informative features, is the better fit of \expmag and \devmag.}
This could suggest that the \modelmag mostly resorted to
a \devmag fit; \rev{adding \devmag colours (again) would not provide any new
information, so the feature selection chooses to add \expmag colours
instead.}
Indeed, comparing the likelihoods of the \devmag and \expmag
    fits\footnote{Available through the \texttt{PhotoObjAll} table in the SDSS
database.}
reveals that the \devmag fit achieved the largest likelihood for $\sim 66\%$ of
the galaxies in the smaller subset.

Returning to Table~\ref{tab:ssfr_results} and now considering experiment~3, which
used the larger subset, but only the four \modelmag colours, we see a
performance similar to experiment~1\rev{, though now with significantly reduced
uncertainties due to the larger sample size}.

Experiment~4 also used the larger subset, but with the eight colours chosen as
the most informative in experiment~2 (the top eight colours in
Fig.~\ref{fig:feature_ranking_ssfr}). As noted, the idea behind this experiment was to
see how features selected on a smaller subset generalise to a larger one. This
is important to know if this method is to be applied to a larger part of SDSS
without any spectroscopically determined \ssfrs to check for consistency with.
\rev{The results from experiment~4 show that the feature selection from
experiment~2 did indeed increase the performance of the method compared to using
the standard colours (experiment~3). The fact that the results of experiment~4
were consistent with those of experiment~2, shows that the most informative
features can indeed be determined from a smaller subset and then used on a
larger. Additionally, it shows that}
%
\knn regression can be an effective
method for determining \ssfrs from photometric data, even when the features
are determined from a much smaller subset.

\begin{figure*}
    \centering
    \subfloat[Template-based model
    predictions.\label{fig:corr:template}]{\includegraphics[width=0.33\textwidth]{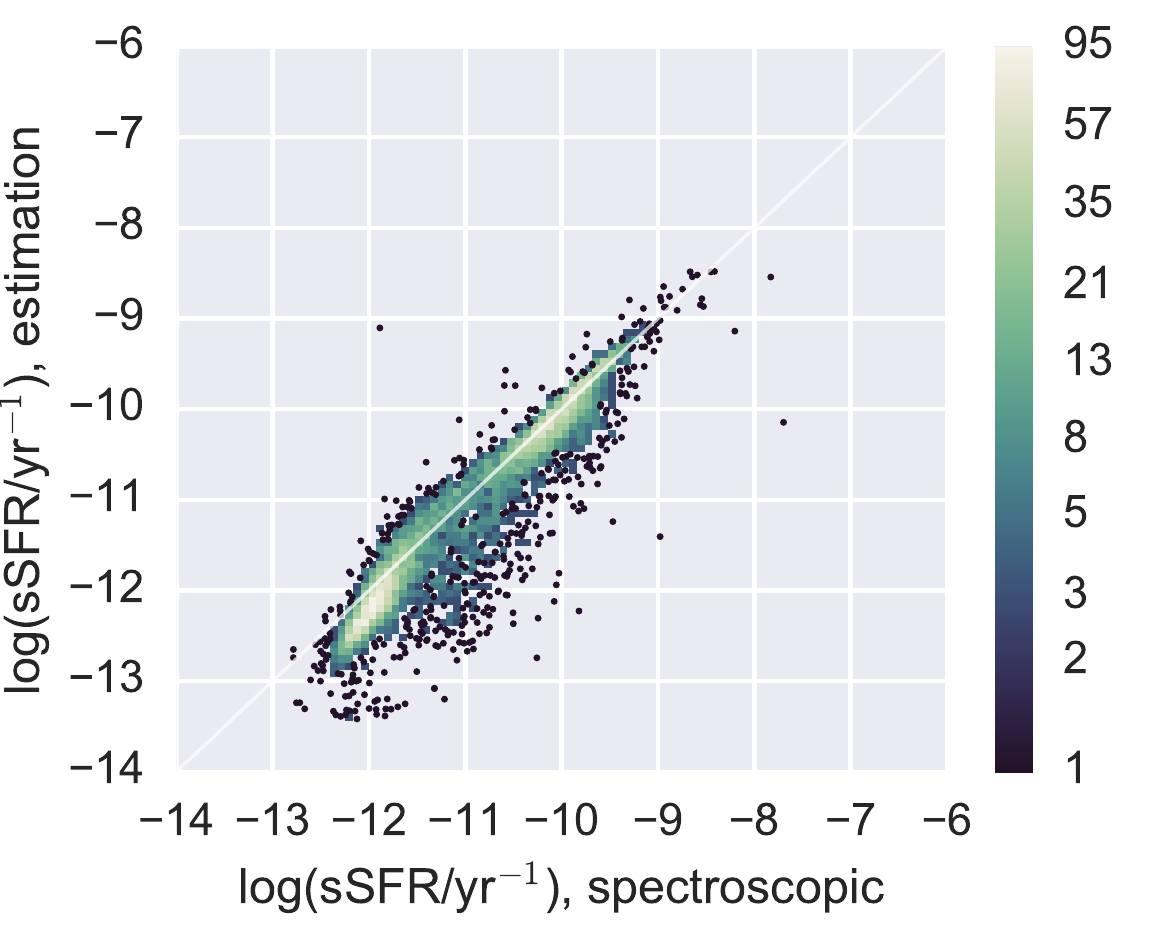}}
    \subfloat[Experiment
    1.\label{fig:corr:e1_ssfr}]{\includegraphics[width=0.33\textwidth]{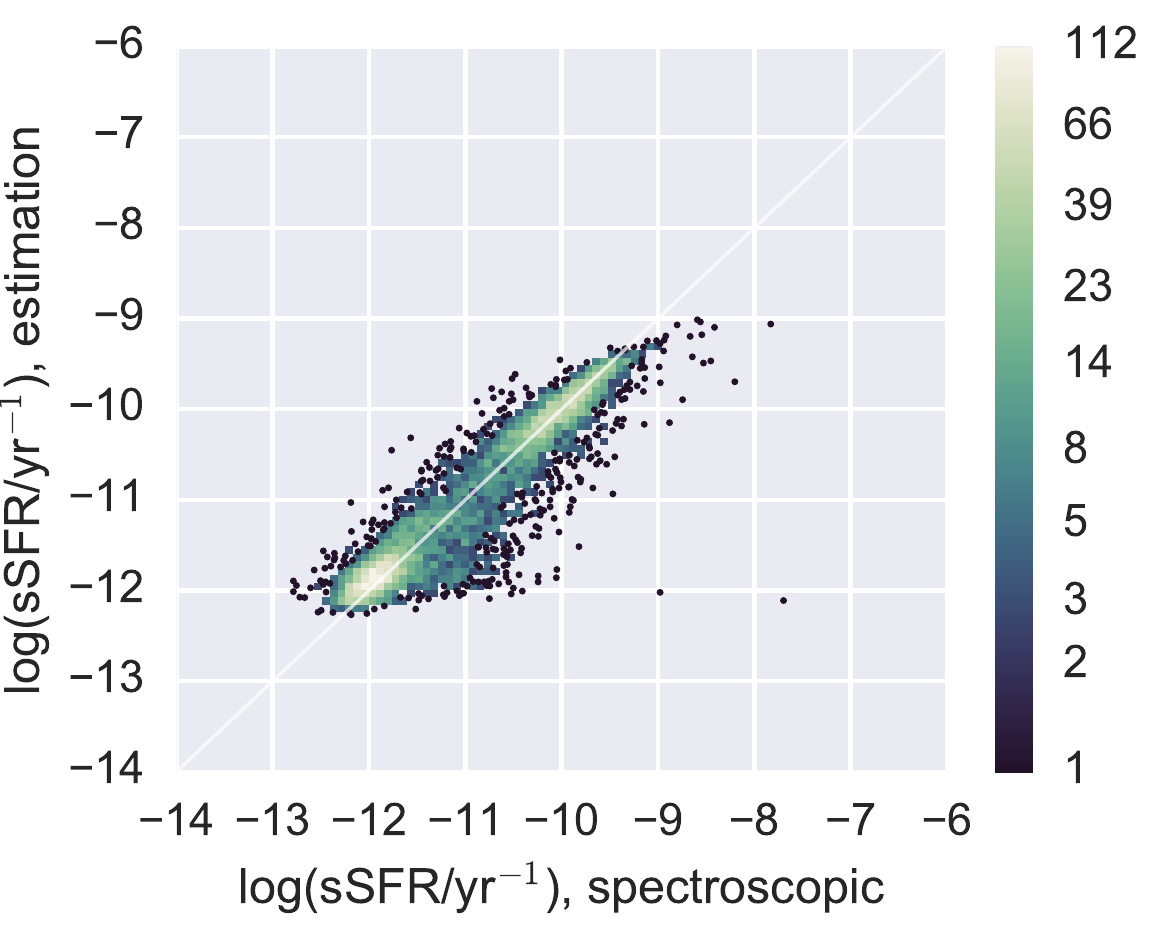}}
    \subfloat[Experiment
    2.\label{fig:corr:e2_ssfr}]{\includegraphics[width=0.33\textwidth]{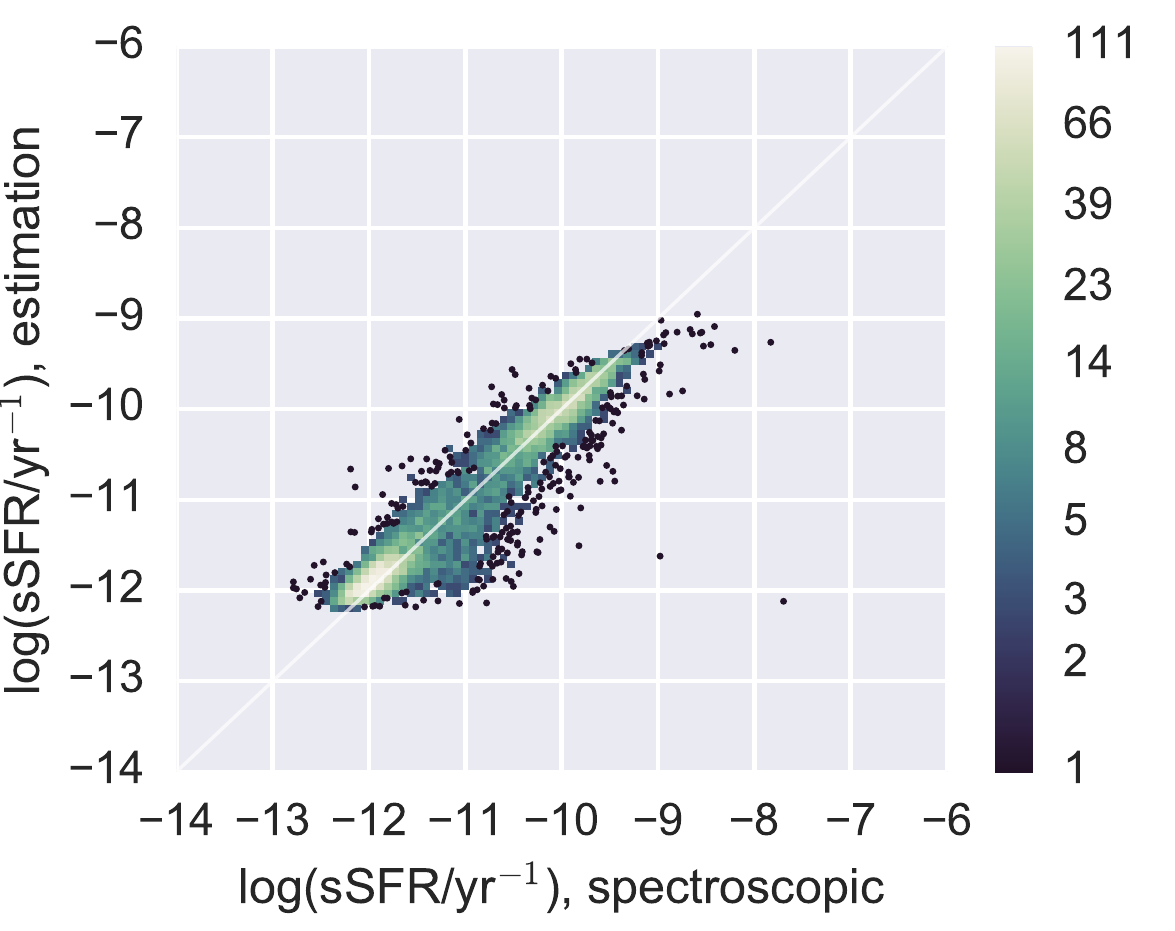}}
    \\
    \subfloat[Experiment
    3.\label{fig:corr:e3_ssfr}]{\includegraphics[width=0.33\textwidth]{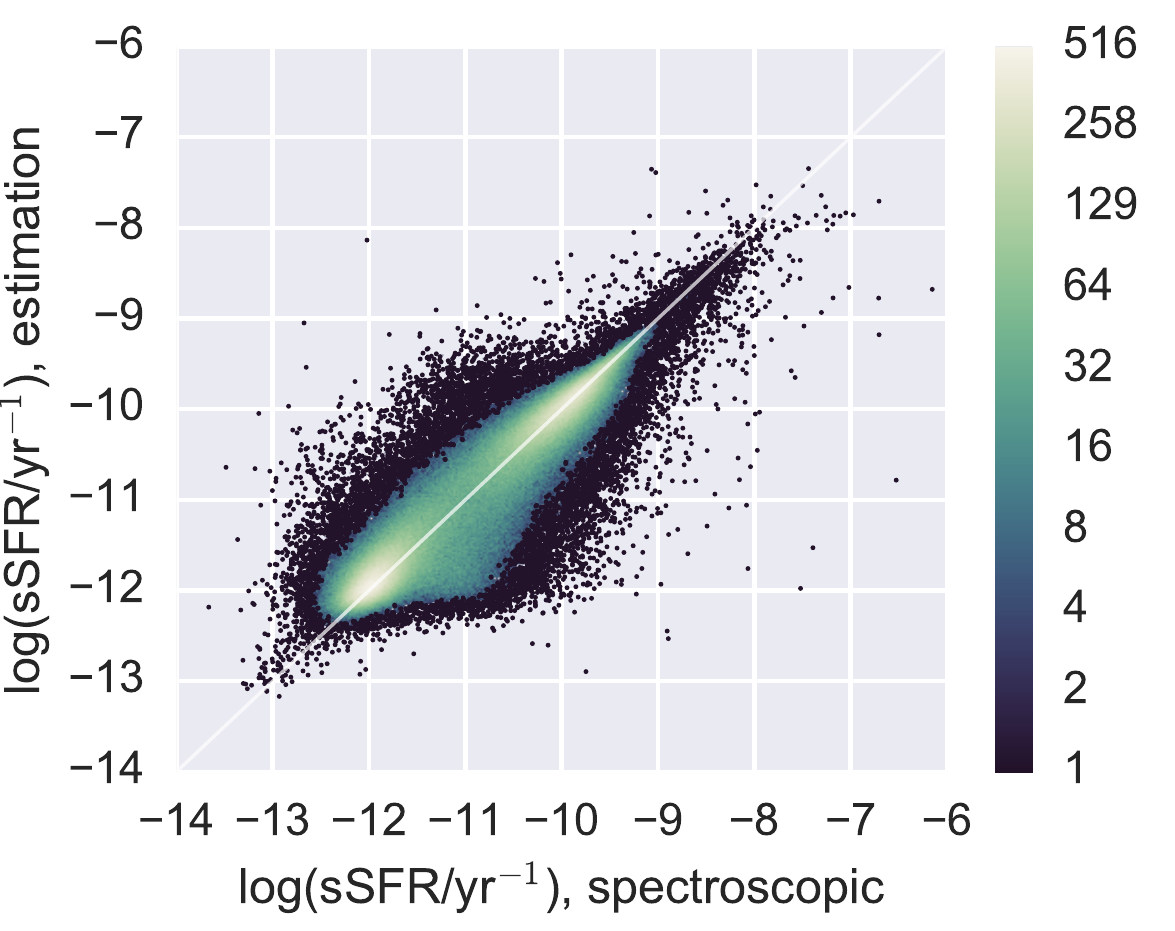}}
    \subfloat[Experiment
    4.\label{fig:corr:e4_ssfr}]{\includegraphics[width=0.33\textwidth]{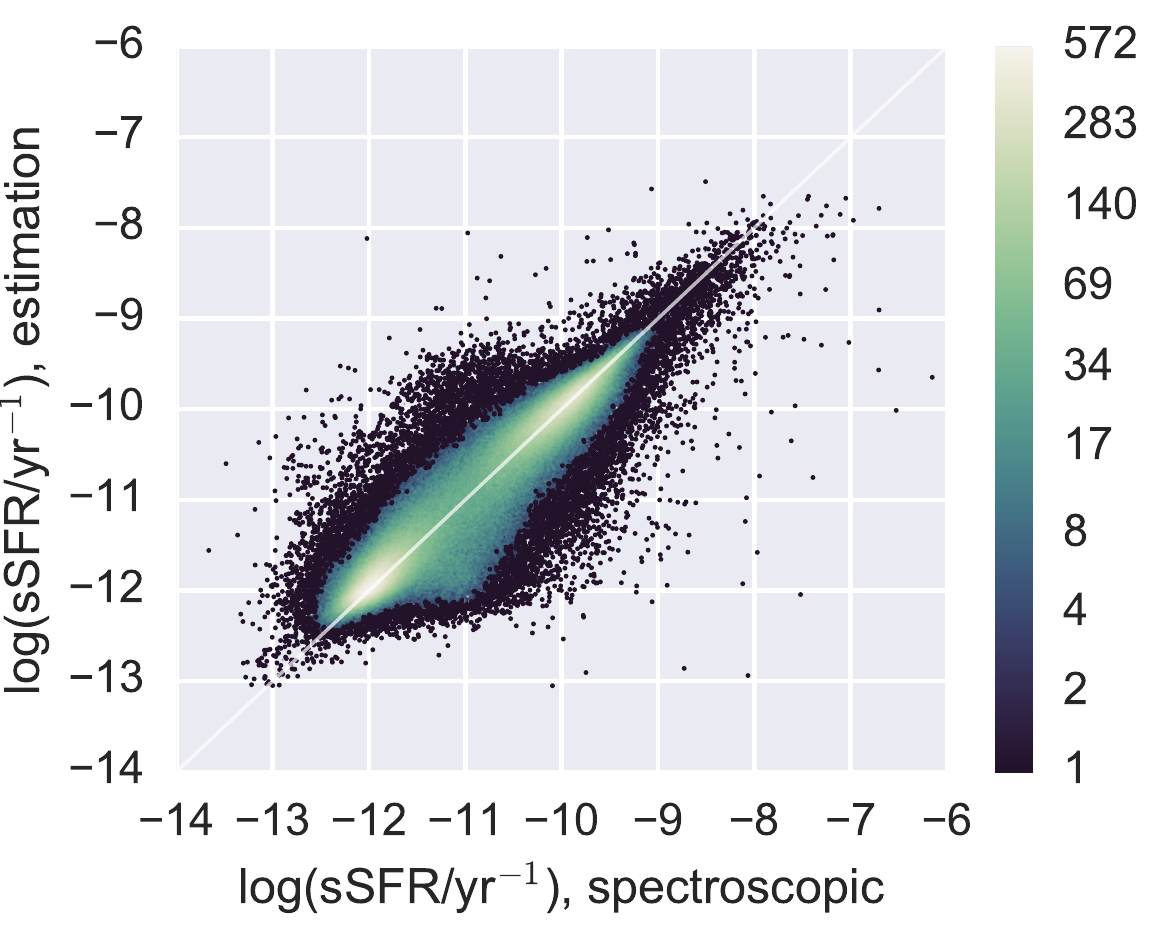}}

    \caption{Correlations between the estimated and spectroscopically determined
        \ssfrs for the template-based model and the four experiments. The colour
        coding indicates the amount of galaxies in each bin.}
    \label{fig:knn_corr_ssfr}
\end{figure*}

Figure~\ref{fig:knn_corr_ssfr} shows the correlations between the spectroscopically
determined \ssfrs and the corresponding estimations from the template-based
model as well as each of the four experiments.
Looking at the estimations from the template-based model
(Fig.~\ref{fig:corr:template}) it is immediately clear where it falls short: it
seems to consistently underestimate the \ssfrs of the low-\ssfr galaxies. The
distribution for high-\ssfr galaxies also seems slightly skewed towards
underestimation.

The estimations done by the \knn regression (Figs.~\ref{fig:corr:e1_ssfr} and
\ref{fig:corr:e2_ssfr}) were clearly better than those from the template-based model.
The distribution for high-\ssfr galaxies seems quite symmetric, while for the
low-\ssfr galaxies it appears slightly skewed towards overestimating the \ssfrs.

The same trends can be seen in the estimations by the \knn regression
on the larger subset (Figs.~\ref{fig:corr:e3_ssfr} and \ref{fig:corr:e4_ssfr});
a symmetric mode for the high-\ssfr galaxies and a slightly skewed
mode for the low-\ssfr galaxies\rev{, though not as pronounced as for the
smaller subset.}

\rev{For all \knn experiments, the distribution at highest \ssfrs seem to skew
towards underestimation. This is likely due to the inherent inability of \knn to
extrapolate beyond the distribution of the training set; as there are only few
data at these \ssfrs, it is likely that the average of the nearest neighbours
(in colour space) will drive the estimated \ssfr towards lower values.}
%
\rev{Apart from choosing a different method than \knn,} an obvious remedy would
be to include more galaxies in the training set to cover more of the
\rev{colour-magnitude} and \ssfr space. Another possibility would be to include
colours from other surveys,
thereby increasing the dimensionality of the \rev{colour-magnitude} space. This could
potentially add the extra information needed in order to move the galaxies
closer to others with similar \ssfrs. Indeed, \citet{Salim2005} showed that a
combination of SDSS and \emph{GALEX} \citep{Martin2005} photometry led to a
significant improvement in the estimation of \sfrs over using just SDSS
photometry. It is natural to assume that this would also be the case with our
method.

When looking at Fig.~\ref{fig:knn_corr_ssfr}, all distributions seem to have a
hump around $(-11,-12)$, where the \ssfrs are somewhat underestimated. It
appears as if galaxies from the green valley get mixed up with quenched
galaxies. The problem also seems to be present for the template-based model,
indicating that there may not be enough information in the SDSS magnitudes to
distinguish these galaxies from quenched ones.
Giving the galaxies a closer look would be an obvious next step to
further increase the accuracy of the methods. It is, however, clear that the
\knn method works equally well for estimating \ssfrs for both main-sequence and
quenched galaxies, which is a rare quality for \ssfr estimation methods in
general.

\begin{figure*}
    \centering
    \subfloat[Template-based model, small subset.\label{fig:res:template_subset}]{
        \includegraphics[width=0.33\textwidth]{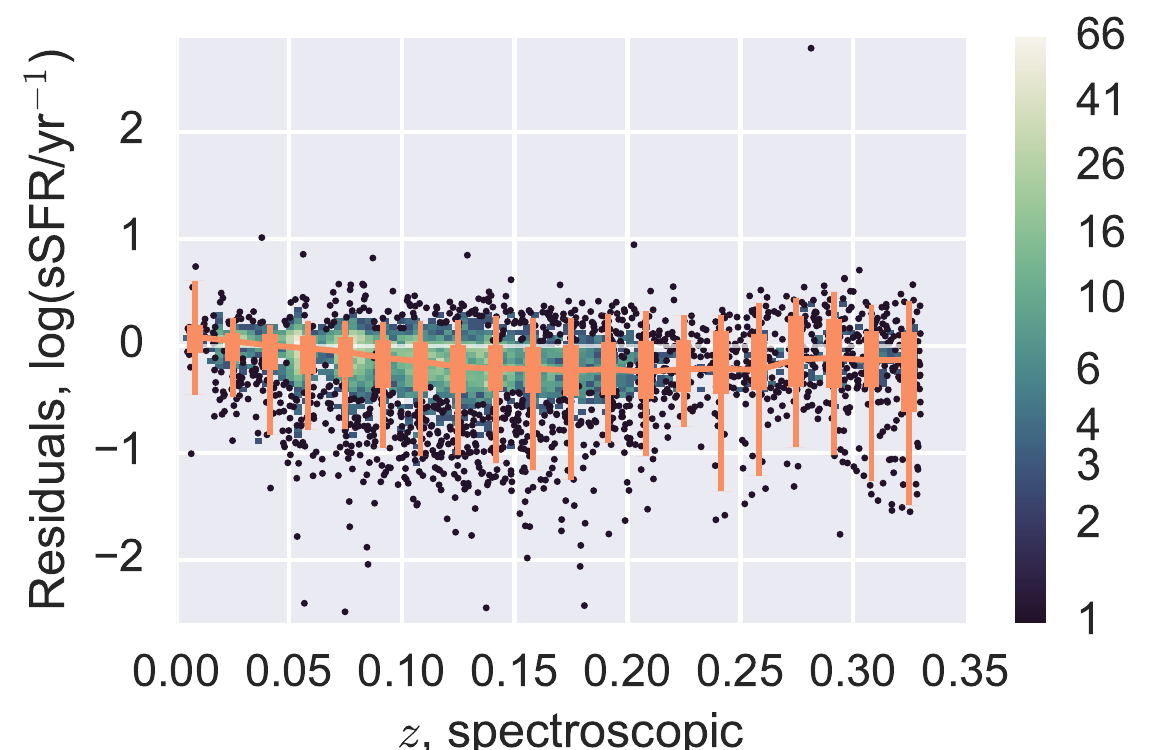}
    }
    \subfloat[Experiment 2.\label{fig:res:subset_ssfr}]{
        \includegraphics[width=0.33\textwidth]{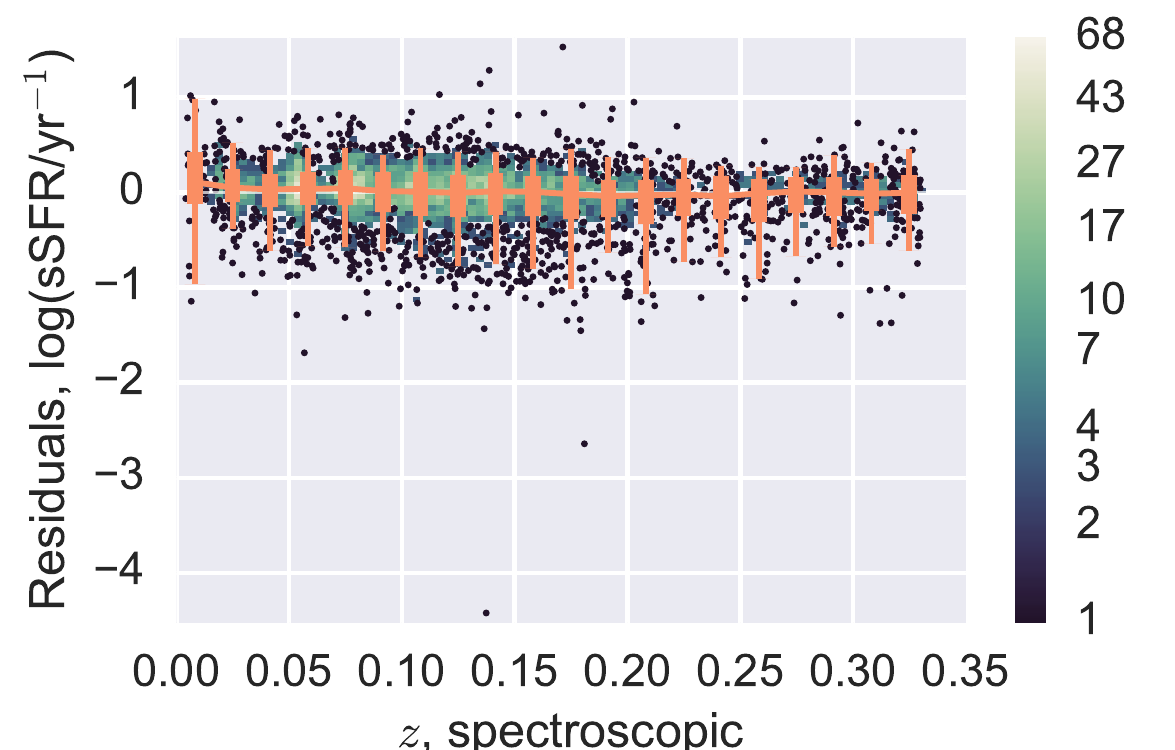}
    }
    %
    \subfloat[Experiment 4.\label{fig:res:bigdata_ssfr}]{
        \includegraphics[width=0.33\textwidth]{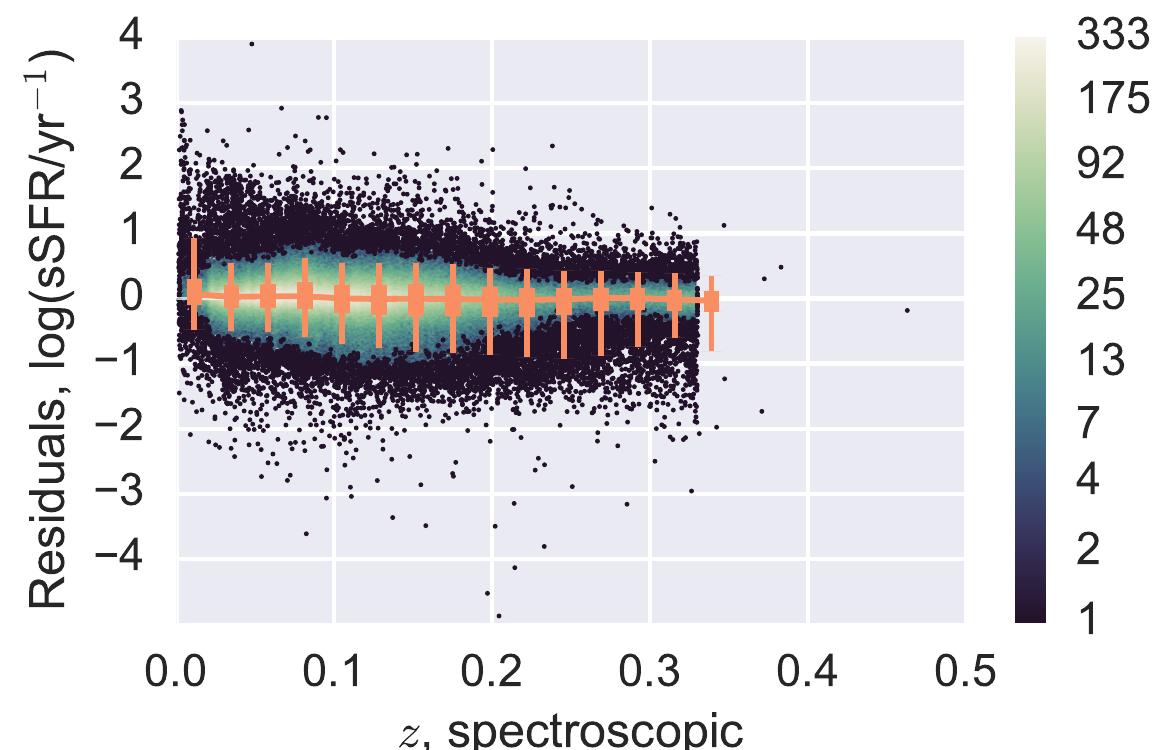}
    }
    \caption{\ssfr residuals as function of redshift for the two galaxy samples
        used in the experiments.  The colour coding of the distributions
        indicates the amount of galaxies in each bin.  The orange line shows the
        running median of the underlying distribution, the thick bars span the
        15.87th through the 84.13th percentile ($\pm 1\sigma$), and the thin
        bars span the 2.28th through the 97.72th percentile ($\pm 2\sigma$).
        Residual plots for experiments~1 and 3 can be found in
        appendix~\ref{app:residuals}.
}
    \label{fig:residuals_ssfr}
\end{figure*}

\rev{Figure~\ref{fig:residuals_ssfr} shows the \ssfr residuals as function of
spectroscopic redshift, with the orange line showing the running median of the
underlying distribution. The thick bars span the 15.87th through the 84.13th
percentile ($\pm 1\sigma$), and the thin bars span the 2.28th through the
97.72th percentile ($\pm 2\sigma$).}

\rev{
The template-based model (Fig.~\ref{fig:res:template_subset}) has a clear
tendency to underestimate the \ssfr throughout the entire redshift range. Our
\knn model (Figs.~\ref{fig:res:subset_ssfr} and \ref{fig:res:bigdata_ssfr})
performs a lot better, with a running median close to $0$ at all
redshifts. The scatter around the running median seems similar for both models,
which is also apparent from Table~\ref{tab:ssfr_results}.

Although the data are limited to rather low redshifts, it is reassuring to see
that there appears to be no significant increase in either bias or scatter, even
at the highest redshifts with our model.}
Note that the redshift was not part of the features used by our method.
Estimation of \ssfrs at all redshifts is based solely on colours and magnitudes
of the galaxies.


\revcolor
\subsection{Redshift experiments}
The accuracy of the \pz experiments is evaluated with the following metrics.
We define the normalised redshift estimation error as $\Delta z' = \Delta z/(1 +
z)$, where $\Delta z = z_\text{phot} - z_\text{spec}$.
Following \citet{Ilbert2006}, we define a catastrophic outlier as a galaxy
with $|\Delta z'| > 0.15$ and $\eta$ as the fraction of catastrophic outliers
in a given experiment. We further use the definition of the normalised median
absolute deviation as $\sigma_\text{NMAD} = 1.48\times\text{median}(|\Delta z'|)$.
Following \citet{Dahlen2013}, we define $\sigma_\text{RMS} =
\langle{\Delta z'}^2\rangle^{1/2}$ and $\sigma_O$ as being the
$\sigma_\text{RMS}$ after catastrophic outliers have been removed. We also
evaluate the bias, given as the mean normalised error, $\biasz = \langle\Delta
z'\rangle$, once again excluding catastrophic outliers.

Table~\ref{tab:results_z} presents the results obtained in the various
experiments. The results are calculated by combining the results from the test
sets in each of the 10 CV folds.
\begin{table*}
    \centering
    \caption{
        Results from the \pz estimation experiments. Evaluation metrics are the
        bias, $\biasz = \langle\Delta z'\rangle$, the normalised
        root-mean-square (RMS) error, $\sigma_\text{RMS} = \langle{\Delta
        z'}^2\rangle^{1/2}$, the RMS error with outliers removed, $\sigma_O$,
        the normalised median absolute deviation, $\sigma_\text{NMAD} =
        1.48\times\text{median}(|\Delta z'|)$, and the fraction of catastrophic
        outliers, $\eta$.
        The standard deviations shown are calculated over the 10 CV folds.
    }
    \label{tab:results_z}
    \sisetup{
        table-alignment = center,
        separate-uncertainty = true,
        fixed-exponent = -2,
        table-omit-exponent = true,
        table-format = +1.2(3),
	}
    \newcommand{\hs}{\hspace{7mm}}
    \begin{threeparttable}
        \begin{tabular}{l @{\hs} c @{\hs}
                S[fixed-exponent=-4, table-figures-uncertainty=4] @{\hs}
                S @{\hs} S @{\hs} S @{\hs}
            S[table-figures-decimal=3]}
            \toprule
            Experiment   & $D$ & {$\biasz / \num{e-4}$} &
            {$\sigma_\text{RMS} / \num{e-2}$} & {$\sigma_O / \num{e-2}$} &
            {$\sigma_\text{NMAD} / \num{e-2}$} & {$\eta / \SI{e-2}{\percent}$} \\
            \midrule

            \multicolumn{7}{c}{SDSS subset of \nsmall galaxies}\\
\midrule
1	&	4	&	-6.13(980)e-04 & 2.19(8)e-02 & 2.17(7)e-02 & 1.72(10)e-02 & 2.56(513)e-02\\
2	&	8\tnote{a}	&	4.29(791)e-04 & 1.83(12)e-02 & 1.82(10)e-02 & 1.45(7)e-02 & 1.28(385)e-02\\
SDSS	&	4\tnote{b}	&	-5.84(1060)e-04 & 2.01(12)e-02 & 1.99(13)e-02 & 1.54(9)e-02 & 3.85(588)e-02\\
\midrule
\multicolumn{7}{c}{SDSS subset of \nlarge galaxies}\\
\midrule
3	&	4	&	3.57(61)e-04 & 2.29(4)e-02 & 2.26(4)e-02 & 1.83(3)e-02 & 3.56(79)e-02\\
4	&	8	&	3.00(56)e-04 & 1.74(3)e-02 & 1.73(3)e-02 & 1.40(2)e-02 & 7.95(232)e-03\\
SDSS	&	4\tnote{b}	&	5.29(538)e-04 & 2.22(2)e-02 & 2.12(2)e-02 & 1.65(2)e-02 & 8.58(89)e-02\\

            \bottomrule
        \end{tabular}
        \footnotesize
        \begin{tablenotes}
            \item[a] Number of features is the median of the ten CV folds.
            \item[b] SDSS additionally fitted a hyperplane in order to make
                estimations.
        \end{tablenotes}
    \end{threeparttable}
\end{table*}

Considering first the experiments on the smaller subset, the SDSS method is
quite consistently outperforming our experiment~1, though the differences are
within one standard deviation. Our experiment~2, however, is consistently
outperforming the SDSS method, though again the differences are within one
standard deviation. Comparing our experiment~1 and 2 shows a much more
significant difference; the chosen features clearly outperformed the four
standard colours.

\begin{figure*}
    \centering
    \includegraphics[width=.9\textwidth]{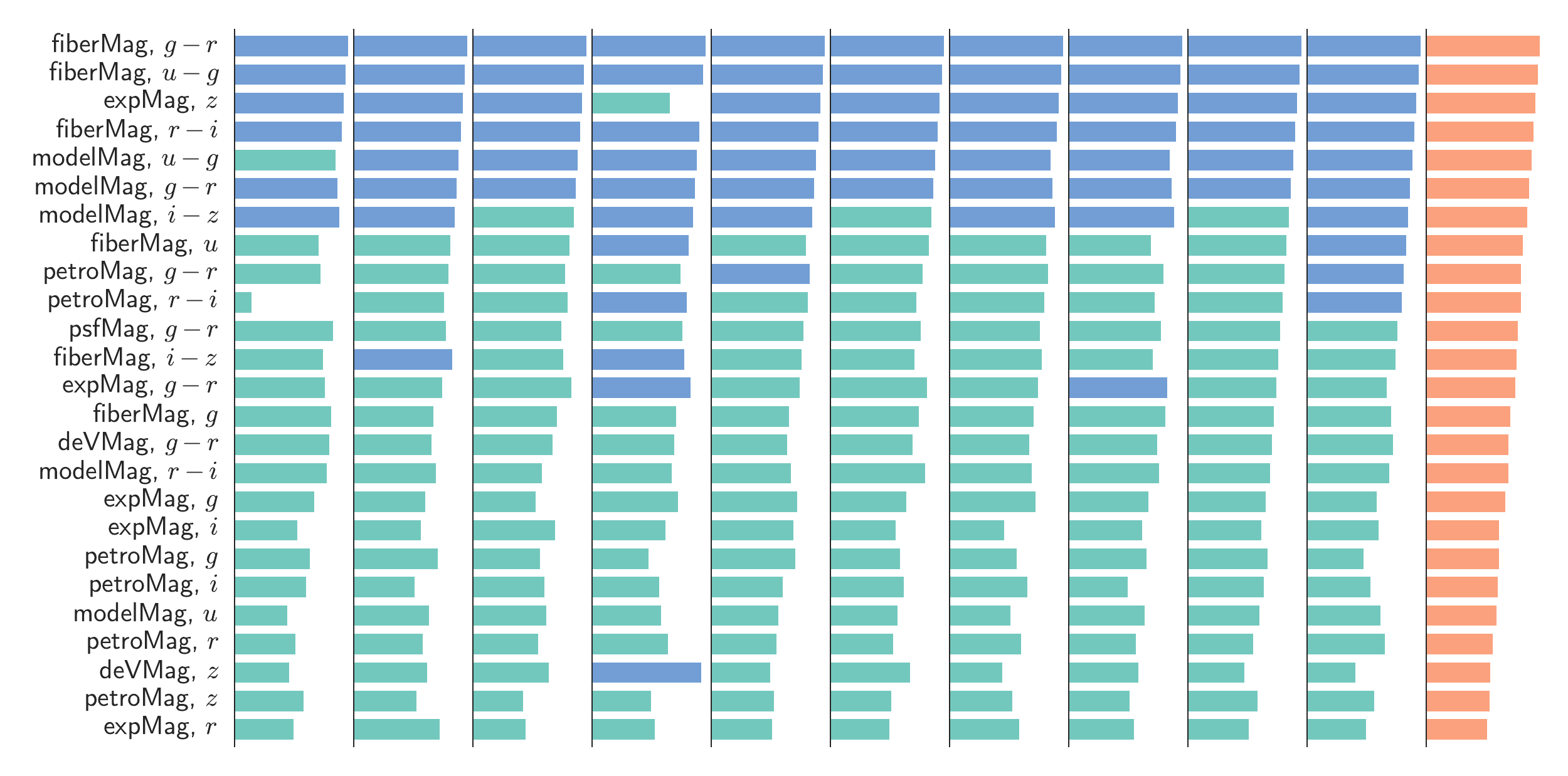}
    \caption{Ranking of the 25 most important features according the feature
        selection in experiment~2.  To the left are the feature names, while the
        rightmost column shows the median rank of each feature across all CV
        folds. Each of the other columns shows the feature ranking in a
        particular CV fold. The larger the bar for a certain feature, the more
        important the feature was. Blue bars show features that were chosen
        during the feature selection as the most informative in a particular CV
        fold. Because of the differences in the data used in each CV fold the
        exact features selected as important, as well as the number of chosen
        features per fold, will vary.  The number of chosen features vary
        from 6 through 11 with a median of $7.5$.
    }
    \label{fig:feature_ranking_z}
\end{figure*}

Figure~\ref{fig:feature_ranking_z} shows the 25 most important features obtained
from the feature selection in experiment~2, with the features chosen as most
informative coloured in blue. The full list of ranked features can
be seen in Fig.~\ref{fig:full_feature_ranking_z}.
The top seven features were quite consistently chosen as the most important,
whereas the remaining chosen features in each CV fold are a lot more scattered
than for \ssfr estimation. The number of chosen features also vary much more:
from six to eleven features are chosen in the folds. The median number of
selected features was $7.5$, so the top eight features in
Fig.~\ref{fig:feature_ranking_z} were chosen as basis for experiment
4. The varying features as well as the number of chosen features for each CV
fold can be an indication that many magnitudes and colours have very similar
information content. Thus, even small differences in the datasets used in each
CV fold can be enough to change the features deemed most informative. Using a
larger dataset for the feature selection will likely make the chosen features
more stable.

It is interesting to see that, while three of the four \modelmag colours are
among the selected features, they are not the most informative. The \fibermag
colours appear to contain more information for \pz estimation.

Another interesting observation is that the $z$ band \expmag was chosen
consistently in all but one CV folds. In the fourth fold, the $z$ band \devmag
was chosen instead of the \expmag. Having a single measure of the $z$ band
magnitude therefore seems to be important for \pz estimation. This is rather
surprising, given the $z$ band's low S/N and the fact that all galaxies in the
small subset have $z \lesssim 0.33$.

Returning again to Table~\ref{tab:results_z}, it is expected that the SDSS
method outperform our experiment~1. Even though we use the same features, the
SDSS estimate uses a hyperplane fit to the nearest 100 samples. This will act as
regularisation, making estimations less susceptible to outliers.

Considering now the experiments on the larger subset, the results are
qualitatively as before, but with significantly reduced error bars. Overall,
SDSS outperforms our experiment~3, which again uses the same features. As
before, this is to be expected. Interestingly, our experiment~3 has a
significantly lower outlier rate $\eta$ than SDSS, but that is likely due to
SDSS training on a larger sample. As \knn is unable to extrapolate beyond just
the mean of the nearest data points, our method will not be able to estimate a
redshift outside the redshift range of the training set. As SDSS has likely used
a much larger training set with more high-$z$ galaxies, it is plausible that their
method has been `confused' by galaxies with similar colours, but much higher
redshifts, leading to a significantly larger redshift estimations, and thus the
possibility of more outliers.

Experiment~4 significantly outperformed both our experiment~3 and, more
interestingly, the \pz estimations from SDSS. We are thus able to achieve much
better performance by using optimal features instead of the standard ones, even
when using additional modelling as SDSS does.

\begin{figure*}
    \centering
    \subfloat[SDSS estimations, small subset.\label{fig:corr:sdss_subset}]{
        \includegraphics[width=0.33\textwidth]{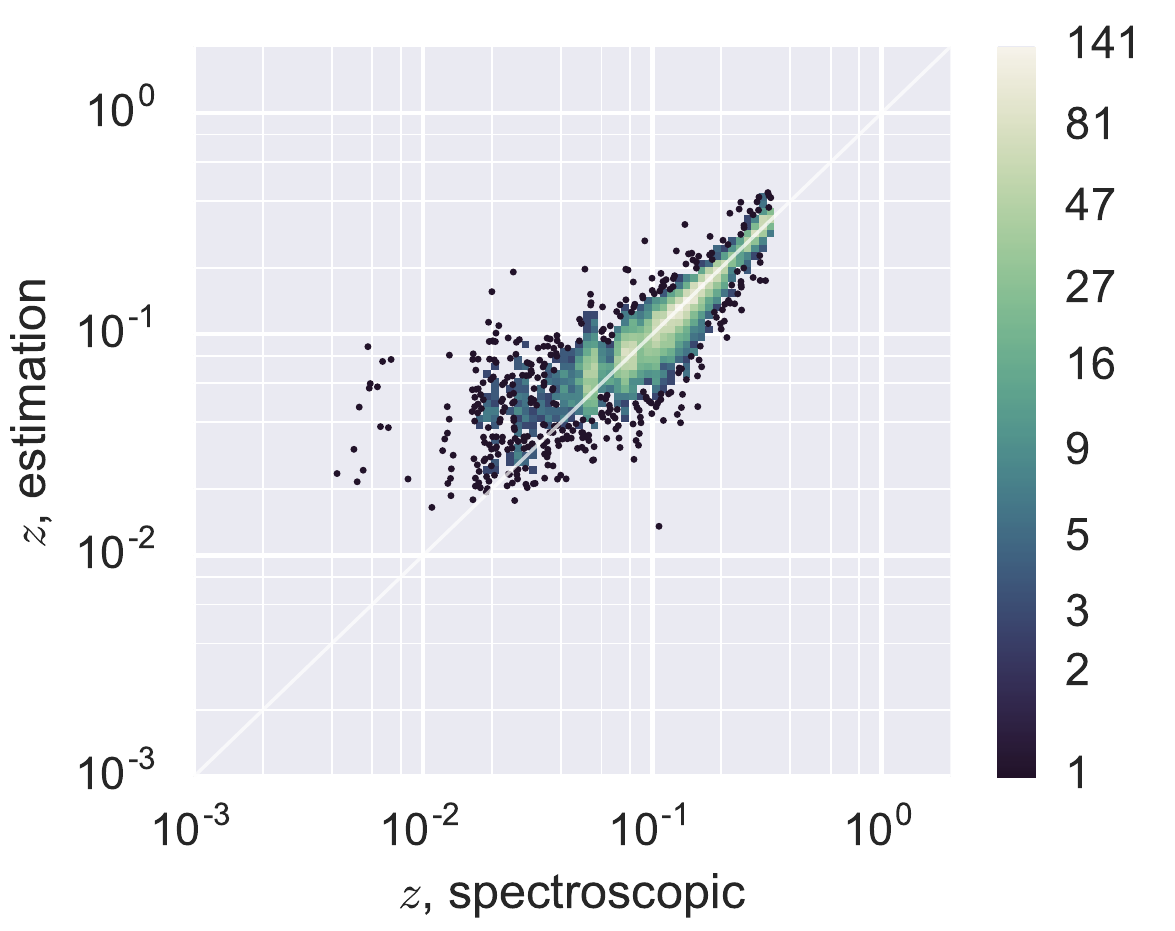}
    }
    \subfloat[Experiment 1.\label{fig:corr:e1_z}]{
        \includegraphics[width=0.33\textwidth]{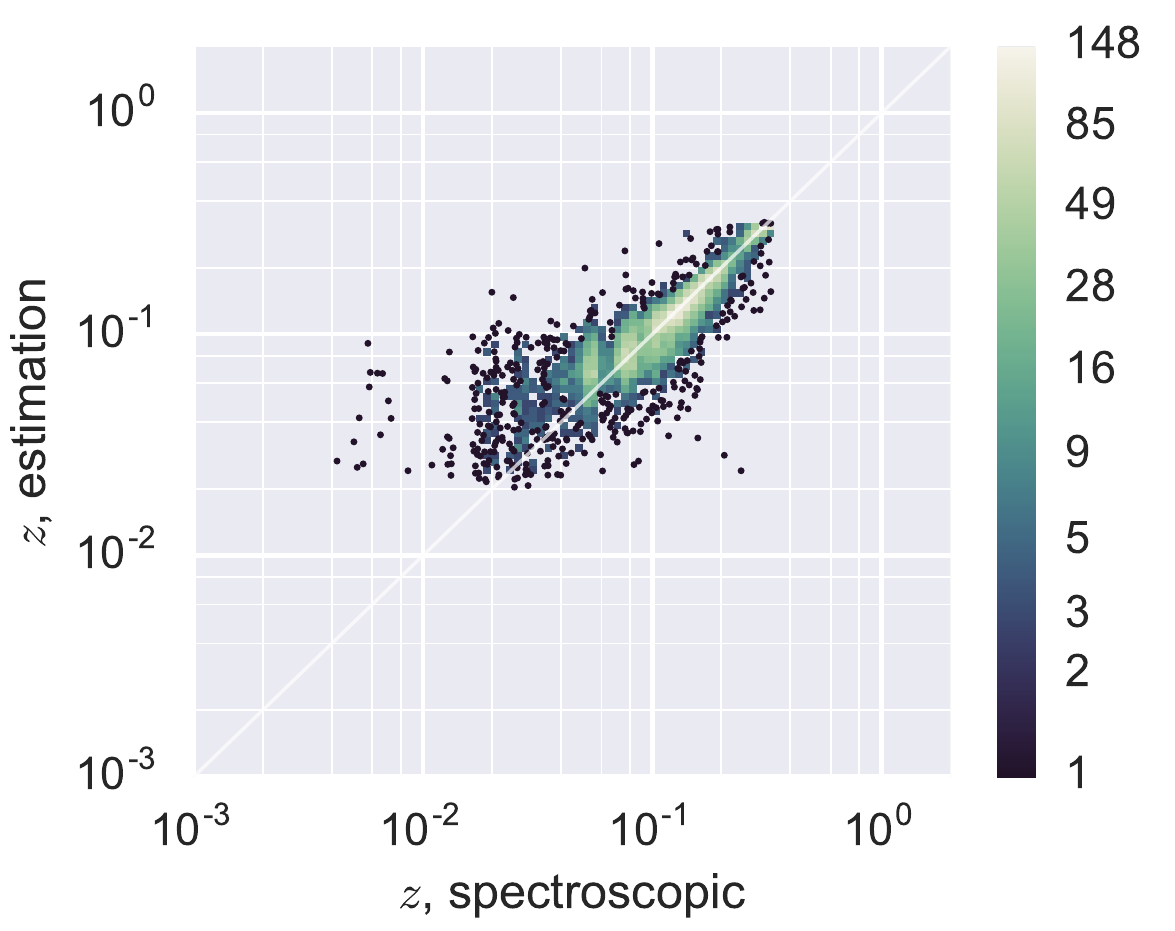}
    }
    \subfloat[Experiment 2.\label{fig:corr:e2_z}]{
        \includegraphics[width=0.33\textwidth]{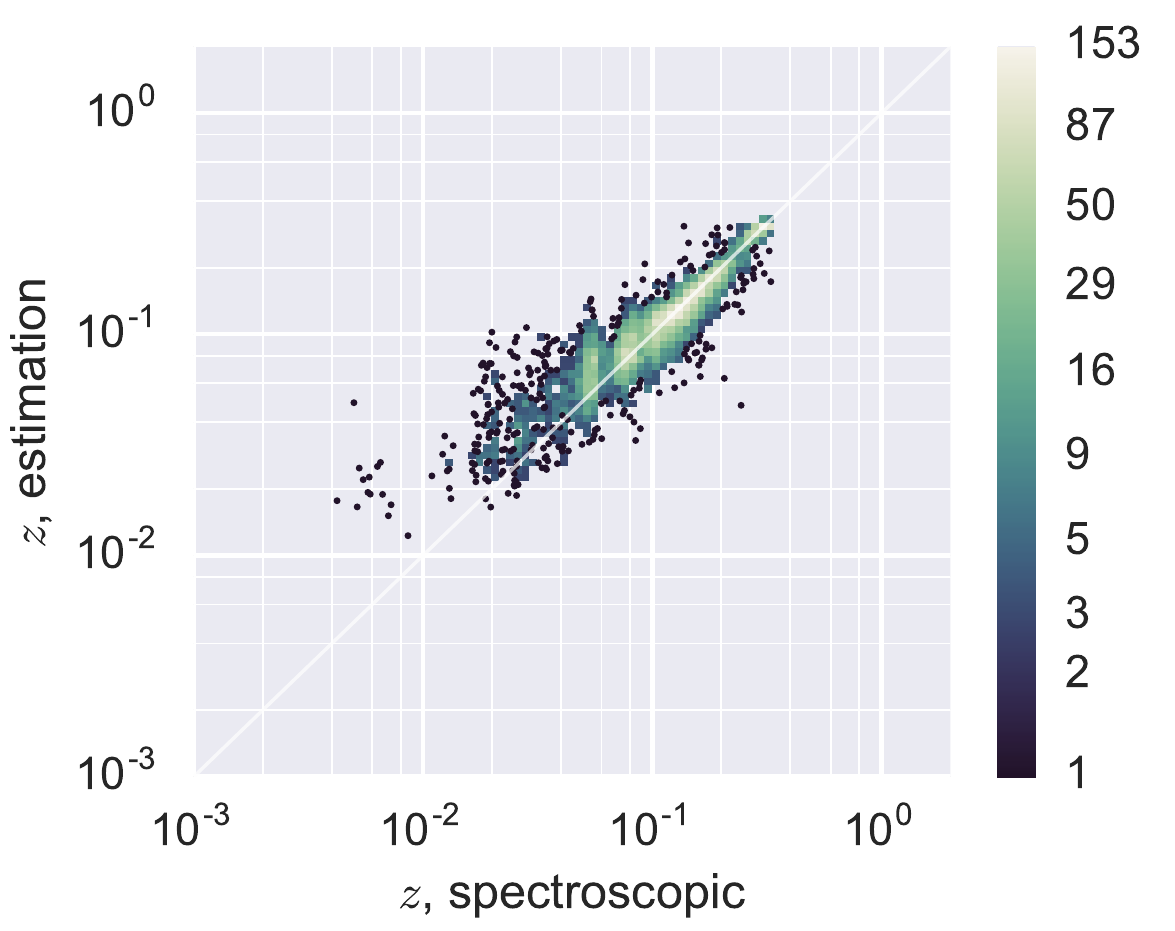}
    }
    \\
    \subfloat[SDSS estimations, large subset.\label{fig:corr:sdss_bigdata}]{
        \includegraphics[width=0.33\textwidth]{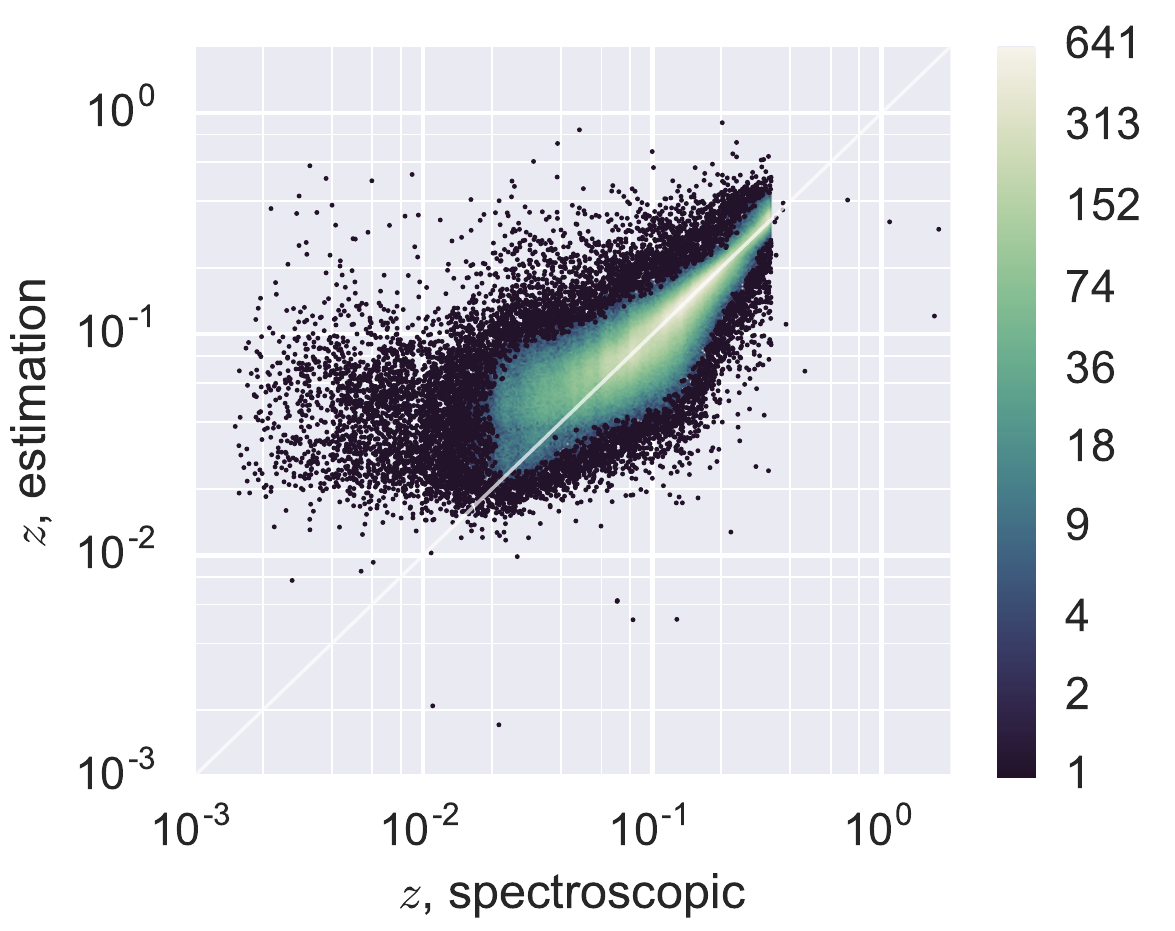}
    }
    \subfloat[Experiment 3.\label{fig:corr:e3_z}]{
        \includegraphics[width=0.33\textwidth]{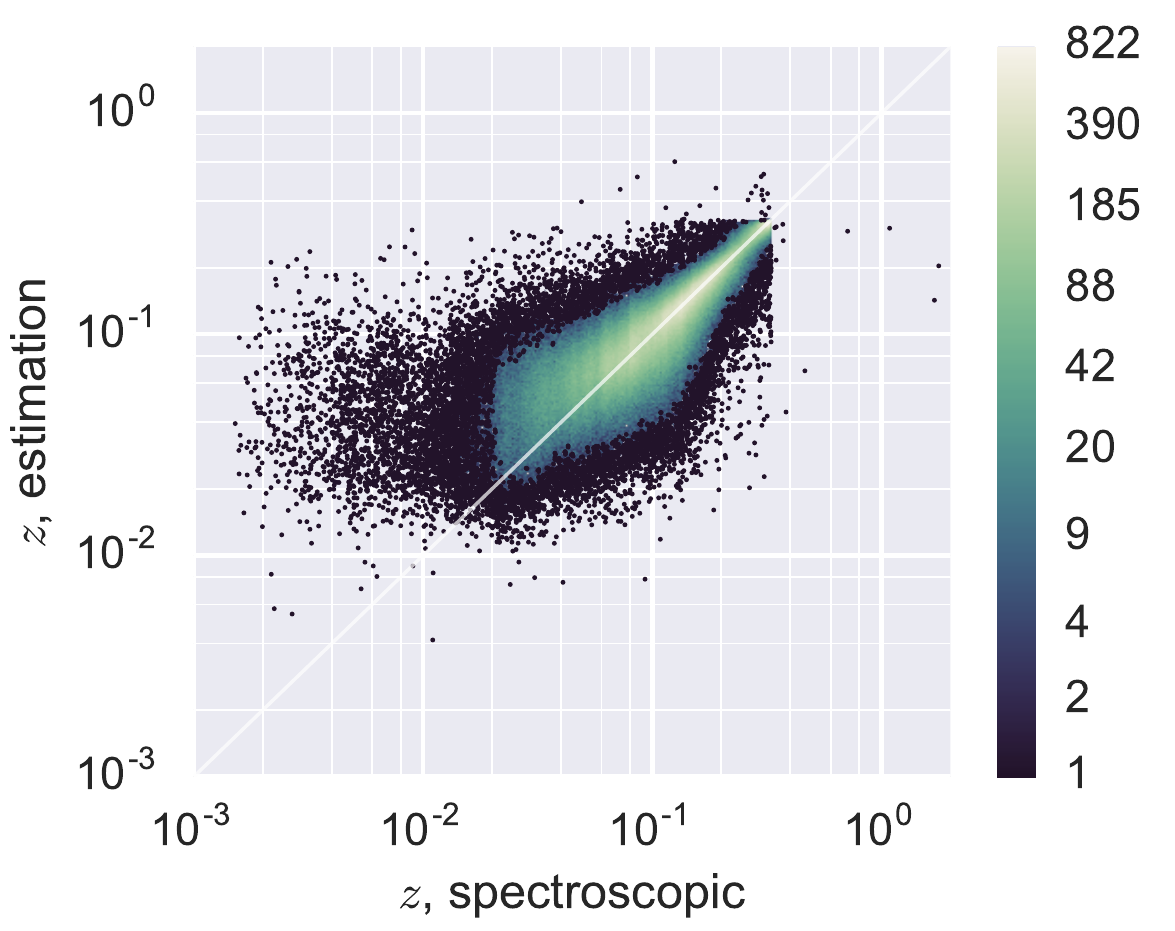}
    }
    \subfloat[Experiment 4.\label{fig:corr:e4_z}]{
        \includegraphics[width=0.33\textwidth]{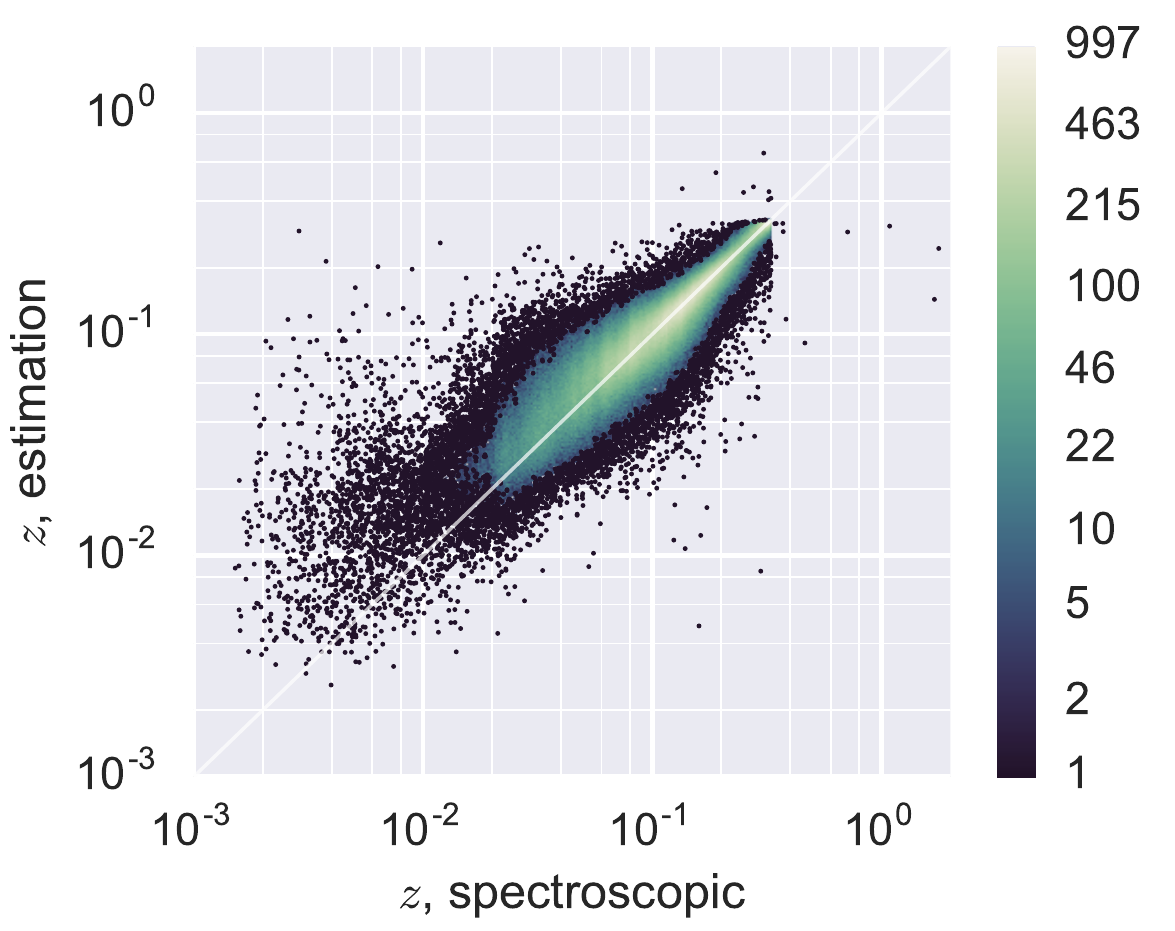}
    }
    \caption{Correlations between the estimated \pz and spectroscopically
    determined $z$ for the SDSS \pz method and our \knn method.  The colour
coding of the distributions indicates the amount of galaxies in each bin.}
    \label{fig:knn_corr_z}
\end{figure*}

Figure~\ref{fig:knn_corr_z} shows correlations between estimated \pz and the
spectroscopically derived redshift.
The spectroscopically determined redshifts have a sharp cut around $z\sim
0.33$, after which there are only few galaxies. This is a result of our data
selection.

Figures~\ref{fig:corr:sdss_subset} and \ref{fig:corr:sdss_bigdata} show the
\pz estimations done by SDSS, \ie not by our model. Figures~\ref{fig:corr:e1_z}
and \ref{fig:corr:e3_z} show the \pz estimations done using our \knn method, but
using only the four \modelmag colours. Finally, Figs.~\ref{fig:corr:e2_z} and
\ref{fig:corr:e4_z} show the \pz estimations done using our \knn method,
including the feature selection.

Focusing first on the experiments using the small subset, we see that
distributions resulting from our \knn
method and SDSS's are, qualitatively, quite similar. Experiment~2, which
used feature selection, seems to have a slightly more symmetric
distribution around the diagonal and appears to work better at the smallest
redshifts, but is otherwise very similar to the other two
experiments. Again, it is worth noting that the SDSS estimations
have likely used a much larger training set than just this subset, which our
model is restricted to. The fact that the results are so similar shows that the
\knn method does not need a large training sample to produce accurate
estimations.

Turning now to the experiments using the large subset, the SDSS method
appears to result in more extreme outliers than ours.
This observation may, however, be misleading.
As the SDSS estimations are likely based on a much larger training set including
galaxies with much higher and lower redshifts, estimations outside the
distribution of our subset are entirely possible. Our
training set is limited to low redshifts, meaning we cannot estimate redshifts
outside this range.
Thus, our estimations have a clear cut at a redshift similar to that of the
spectroscopic redshift cut, with only a few outliers due to a few high redshift
galaxies.

Ignoring for a moment estimations above $z_\text{phot} \sim 0.33$, the SDSS
estimation seems to perform better than our experiment~3
(Fig.~\ref{fig:corr:e3_z}), which only used the four \modelmag colours. The
SDSS \pz distribution seems tighter around the diagonal, which is likely a
result of the hyperplane fit acting as regularisation. Both methods do, however,
significantly overestimate at the lowest redshifts.

Figure~\ref{fig:corr:e4_z} shows the estimations done by our \knn method, using
the features obtained from the feature selection process in experiment~2.
Compared to Fig.~\ref{fig:corr:e3_z}, there is less scatter and the distribution
is significantly tighter around the diagonal. Comparing with the SDSS
estimations (Fig.~\ref{fig:corr:sdss_bigdata}), we perform significantly better
at the lowest redshifts,
with the added bonus of a more symmetric distribution.

\begin{figure*}
    \centering
    \subfloat[SDSS \pz, small subset.\label{fig:res:sdss_subset}]{
        \includegraphics[width=0.45\textwidth]{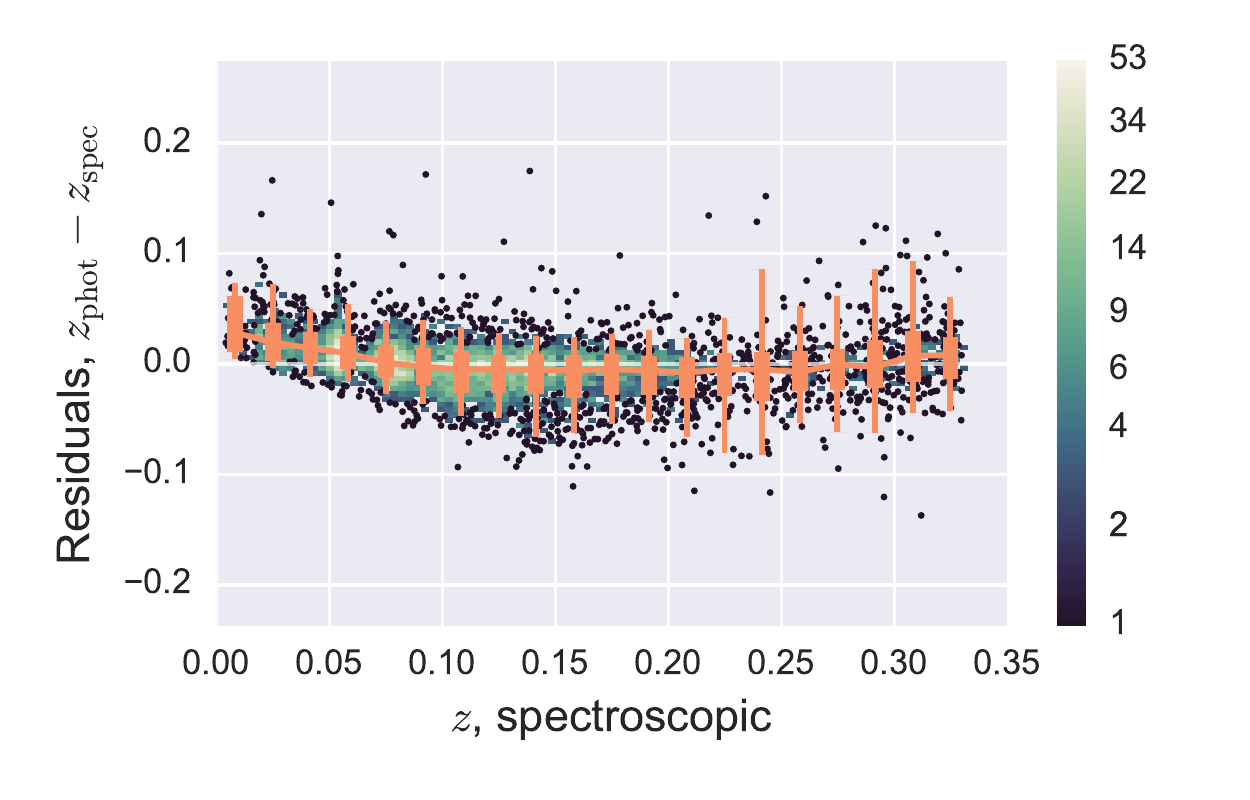}
    }
    \subfloat[SDSS \pz, large subset.\label{fig:res:sdss_bigdata}]{
        \includegraphics[width=0.45\textwidth]{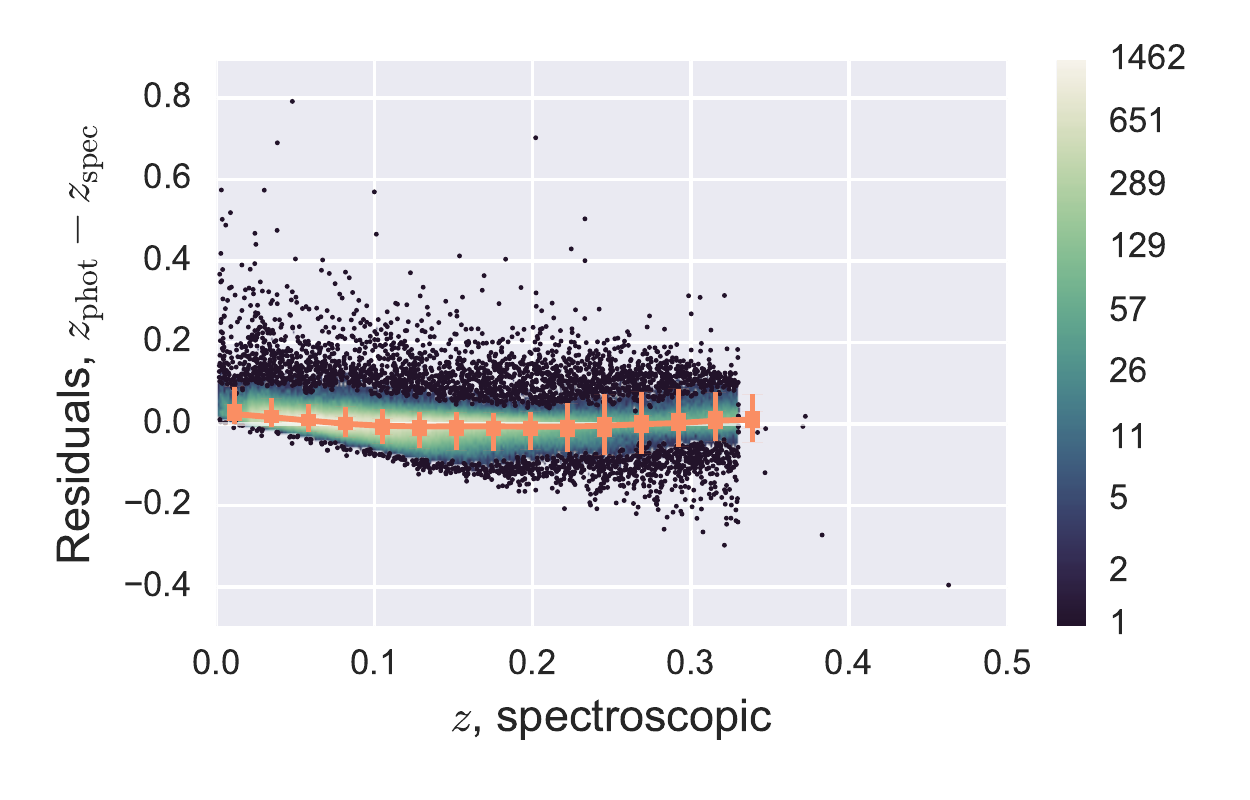}
    }
    \\
    \subfloat[Experiment 2.\label{fig:res:subset}]{
        \includegraphics[width=0.45\textwidth]{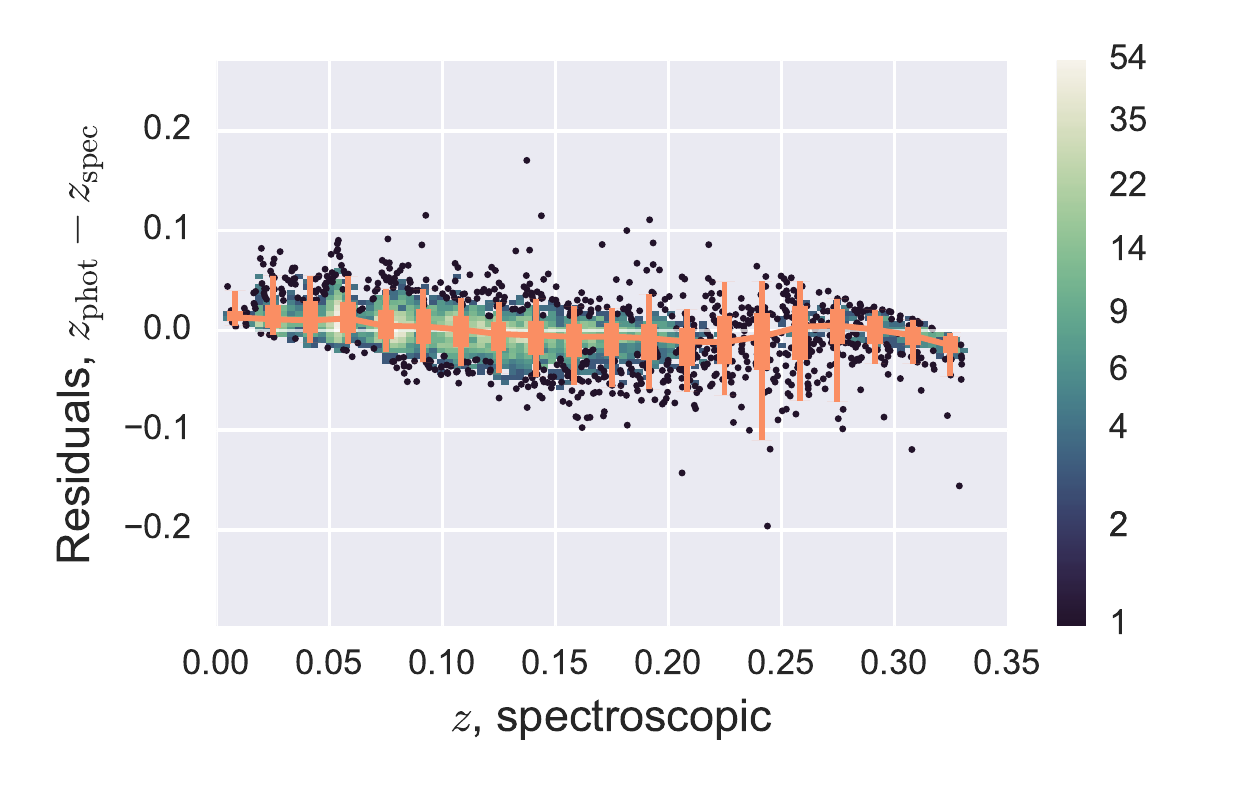}
    }
    \subfloat[Experiment 4.\label{fig:res:bigdata}]{
        \includegraphics[width=0.45\textwidth]{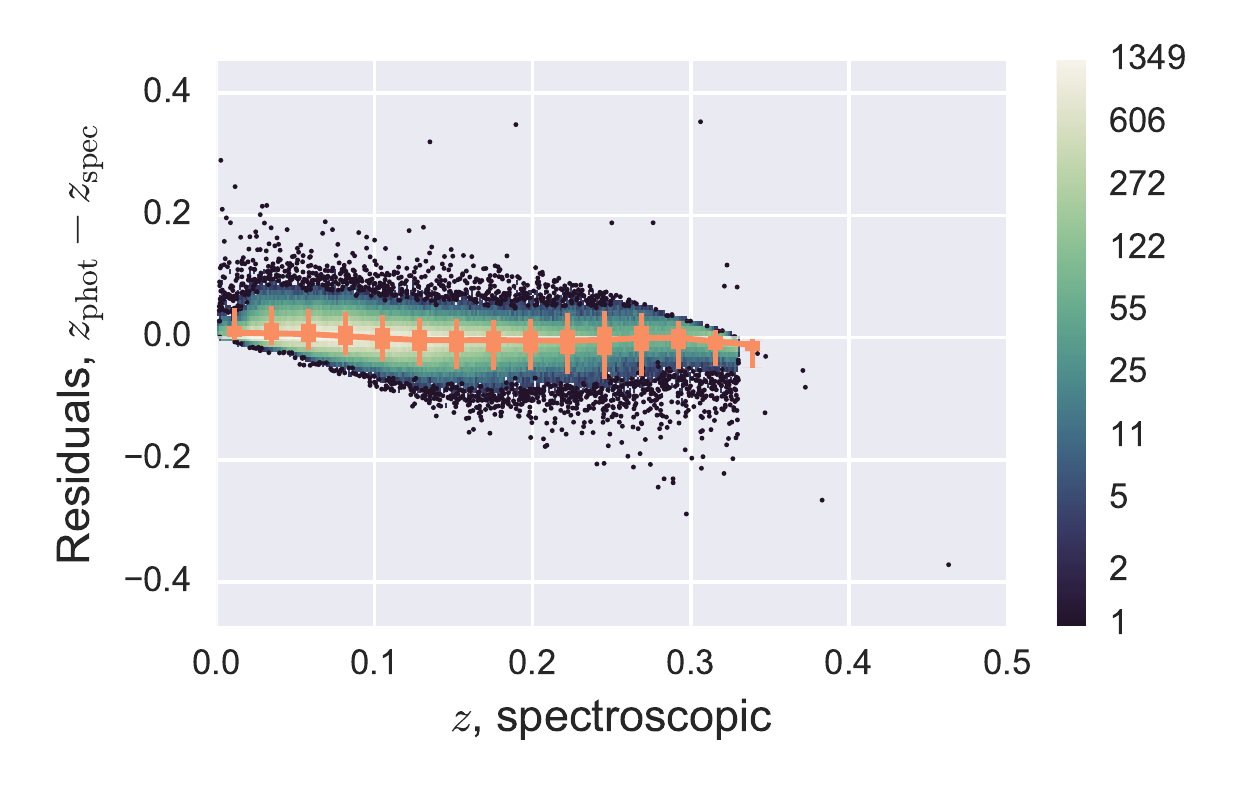}
    }
    \caption{Redshift residuals as function of redshift for the two galaxy
        samples used in the experiments.  The colour coding of the distributions
        indicates the amount of galaxies in each bin.  The orange line shows the
        running median of the underlying distribution, the thick bars span the
        15.87th through the 84.13th percentile ($\pm 1\sigma$), and the thin
        bars span the 2.28th through the 97.72th percentile ($\pm 2\sigma$).
        The sharp slopes seen in \protect\subref{fig:res:subset} and
        \protect\subref{fig:res:bigdata} are a consequence of the training set
        containing only few galaxies with $z \gtrsim 0.33$. As the \knn method
        is not well suited for extrapolation, only few galaxies will have an
    estimated \pz$\gtrsim 0.33$. Residual plots for experiments~1 and 3 can be
found in appendix~\ref{app:residuals}.}
    \label{fig:residuals_z}
\end{figure*}

Finally, Fig.~\ref{fig:residuals_z} shows the \pz residuals as function of
spectroscopic redshift.
The rather sharp slopes at $z \sim 0.3$ for our estimations are a
result of the cut in the spectroscopic redshift as discussed previously. The
SDSS estimations do not exhibit this slope due to the larger training set,
making extrapolation beyond $z \sim 0.3$ much more likely.

Considering the \pz experiments on the small subset, there is not much difference
between the estimates from SDSS (Fig.~\ref{fig:res:sdss_subset}) and those from
our \knn method (Fig.~\ref{fig:res:subset}). Note that the estimations from our
method have been obtained using feature selection.
Both estimation methods appear to overestimate the redshift at low redshifts,
though it is more pronounced for the SDSS method. At higher redshifts, the
methods both slightly underestimate the redshifts.
At the highest redshifts the SDSS method appears to overestimate slightly, while
our \knn method seems to underestimate the redshift. This underestimation is a
consequence of the slope, as the training set used for our method contains only
a few galaxies with $z \gtrsim 0.33$. Therefore, one should not conclude too much
from this underestimation.

The picture is very similar when considering the experiments on the large
subset (Figs.~\ref{fig:res:sdss_bigdata} and \ref{fig:res:bigdata}). There is a
tendency to overestimate the redshift at small $z$ and underestimate it at
higher $z$. For both experiments, however, the median residual is always close
to zero.
There are a few extra galaxies at $z > 0.5$ not shown in these plots, in order
to keep the main galaxy sample detailed. Both methods significantly
underestimate the redshifts of these high-$z$ galaxies with roughly the same
amount.

From the plots in Fig.~\ref{fig:residuals_z}, it is clear that
using just the most important features, we can achieve a similar
performance to fitting a hyperplane to the nearest neighbours, though at a
much lower computational cost once the features have been determined.

\color{black}

\section{Discussion and Conclusions}
\label{sec:conclusion}

In the coming years, increasingly larger astronomical surveys will produce
unprecedented amounts of data. Many of these data will require accurate
estimations in near real-time, which is not feasible with traditional methods.
Machine learning is well-suited to address this challenge.

This work has exemplified this by \rev{showing how machine learning can be used
to not only estimate specific star formation rates (\ssfrs) and photometric
redshifts (\pzs) of galaxies, but also to identify the most informative features
for these tasks, thereby increasing accuracy further. We have shown how}
the simple, yet powerful non-parametric $k$ nearest
neighbours (\knn) method significantly outperforms the traditional
method of simulated template spectra for estimating \ssfrs, achieving a RMSE of
\tssfrrmsea
(the $\pm$ values refer to the standard deviation over the
  non-independent CV folds)
compared to a template-based method's \tssfrrmsec, when using the exact same
input features. Adding a \emph{feature selection} to the \knn method increased
its performance, achieving a RMSE of \tssfrrmseb. \rev{Similarly, the fraction
of catastrophic outliers reduced from the template-based method's \tssfretac to
\tssfretab, when using \knn and feature selection.}

\rev{We see a similar pattern when considering \pz estimation. Here, the
    \knn
method achieves a normalised
median absolute deviation of \tzsignmada, which reduces to \tzsignmadb when
doing feature selection, compared to \tzsignmadc achieved by SDSS. The method
used by SDSS even included a hyperplane fit and while that improves estimations,
it also significantly increases the required amount of computations per
estimate.}

Applying the \knn method to a larger subset of SDSS of \nlarge galaxies, we
achieve a RMSE of \bssfrrmsea for \ssfr estimation, when using the same four features as the
template-based method. By using the features selected in the feature selection
on the smaller subset, we are able to decrease the error further to \bssfrrmseb.
\rev{For \pz estimation, we achieve
a normalised median absolute deviation of \bzsignmada,
which reduces to \bzsignmadb when doing feature selection, compared to
\bzsignmadc achieved by SDSS.
This shows that not only can features selected for a smaller subset be directly
transferred to a much larger one yielding similar performance, the estimations
done by the selected features can even significantly outperform more
computationally intensive modelling.
}

An advantage of a template-based method is the gain in physical knowledge from
the simulations. The feature selection for the \knn method can provide hints to
which features contain the most information, but a deeper understanding of why
these particular features contain more information requires further
investigation and is outside the scope of this work.
The \knn method does, however, have advantages over a template-based method in
that it
is faster and
will not be prone to \rev{errors resulting from} approximations or wrong
assumptions done in the model building process.
This study shows that \rev{machine learning methods, here exemplified by \knn
regression,} should be considered a viable alternative to the traditional
template-based method in situations where high accuracy or computational
efficiency is required. \rev{In particular, adding a feature selection step to
the machine learning methods, instead of relying on traditionally used features,
should be considered part of the standard toolbox.}

\section*{Acknowledgements}
\label{sec:acknowledgements}
We sincerely thank Jarle Brinchmann for \rev{providing us with} photometric
estimations of masses and star formation rates \rev{(through private
communication), and for the spectroscopic \sfrs made available at
\url{http://wwwmpa.mpa-garching.mpg.de/SDSS/DR7/sfrs.html}}. We also thank the
SDSS collaboration for making their reduced data available.

This research made use of NASA's Astrophysics Data System;
NumPy \& SciPy \citep{Jones2001,Walt2011};
the IPython package
\citep{Perez2007};
Scikit-learn \citep{Pedregosa2011};
Pandas \citep{McKinney2010};
matplotlib, a Python library for publication quality graphics
\citep{Hunter2007};
Seaborn \citep{Waskom2016}.

KSS, CI, and KSP  gratefully acknowledge support from The Danish Council for
Independent Research $|$ Natural Sciences through the project ``Surveying the
sky using machine learning''.

Funding for the SDSS and SDSS-II has been provided by the Alfred P. Sloan
Foundation, the Participating Institutions, the National Science Foundation, the
U.S. Department of Energy, the National Aeronautics and Space Administration,
the Japanese Monbukagakusho, the Max Planck Society, and the Higher Education
Funding Council for England. The SDSS Web Site is \url{http://www.sdss.org/}.

The SDSS is managed by the Astrophysical Research Consortium for the
Participating Institutions. The Participating Institutions are the American
Museum of Natural History, Astrophysical Institute Potsdam, University of Basel,
University of Cambridge, Case Western Reserve University, University of Chicago,
Drexel University, Fermilab, the Institute for Advanced Study, the Japan
Participation Group, Johns Hopkins University, the Joint Institute for Nuclear
Astrophysics, the Kavli Institute for Particle Astrophysics and Cosmology, the
Korean Scientist Group, the Chinese Academy of Sciences (LAMOST), Los Alamos
National Laboratory, the Max-Planck-Institute for Astronomy (MPIA), the
Max-Planck-Institute for Astrophysics (MPA), New Mexico State University, Ohio
State University, University of Pittsburgh, University of Portsmouth, Princeton
University, the United States Naval Observatory, and the University of
Washington.

\bibliography{refs}

\bibliographystyle{mnras}

\appendix
\section{Massively parallel greedy feature selection}
\label{app:feature_selection}

While greedy procedures \rev{such as forward or backward feature selection} are
significantly faster than the exhaustive search for the best-performing
features, they can still be very time-consuming, even on training sets of
moderate sizes. One way to accelerate such a feature selection step is to speed
up the involved nearest neighbour computations. In the literature, various
techniques can be found for this task. Typical methods are
\emph{\kdtrees}~\citep{Bentley1975} or \emph{locality-sensitive
hashing}~\citep{Indyk1998}. However, such tools either perform poorly in higher
dimensions or only yield approximate answers. A recent trend in data analytics
is to resort to (exact) parallel implementations for many-core devices such as
today's \emph{graphics processing units} (\GPUs). For instance,
\cite{Garcia2010} make use of highly-tuned \GPU~matrix multiplication libraries
for nearest neighbour search. Other schemes are based on, \eg adapted spatial
search structures~\citep{Cayton2012,GiesekeHOI2014,Nakasato2012}.

For the work at hand, we make use of a massively-parallel matrix-based
implementation that addresses incremental feature selection and nearest neighbour
models
recently proposed by \cite{GiesekePOI2014}. For the sake
of completeness, we briefly outline the general workflow of the implementation:
The general workflow for the case of forward selection is sketched in
Algorithm~\ref{alg:fastparallelbackwardselection}. For a given training set $S$
of labelled samples, start with an empty distance matrix $\mathbf{M} \in
\Reals^{\tsize \times \tsize}$ that contains the current distances between all
pairs of training samples. Further, the array \texttt{selected\_dimensions}
indicating the selected features and the array \texttt{val\_errors} are
initialized. The forward feature selection process starts in Step~$4$: The
procedure \textsc{GetValidationErrors} computes, for each dimension $j$ that has
not yet been selected (\ie \texttt{selected\_dimensions}[$j$]=0), the
cross-validation error for the case of dimension $j$ being `added' to the
current set of features. These values are stored in the array
$\texttt{val\_errors}$ and the procedure \textsc{GetMinDim} returns the index of
the smallest error contained in it (thus, $i_{\min}$ corresponds to the
dimension whose addition leads to the smallest cross-validation error).
Afterwards, both \texttt{selected\_dimensions} and $\textbf{M}$ are updated accordingly, where $\textbf{M}^{i_{\min}}$ denotes the all-pairs distance matrix based on dimension $i_{\min}$ only.

\begin{algorithm}[t]
\caption{\textsc{ForwardSelection($S$, $\numfeat$)}}
\label{alg:fastparallelbackwardselection}
\small
\begin{algorithmic}[1]
\REQUIRE Training set $S=\{(\vec{x}_1,y_1),\ldots,(\vec{x}_\tsize,y_\tsize)\} \subset \Reals^\tdim \times \Reals$ and a number $\numfeat < \tdim$ of desired features.
\ENSURE Array \texttt{selected\_dimensions} with selected features.
\STATE Initialize empty distance matrix $\textbf{M} \in \Reals^{\tsize \times \tsize}$;
\STATE \texttt{int selected\_dimensions}[$\tdim$] = $\{0,\ldots,0\}$;
\STATE \texttt{float val\_errors[}$\tdim$\texttt{]};
\FOR{$i=1, \ldots, \numfeat$}
\STATE  \texttt{val\_errors} = \textsc{GetValidationErrors($\textbf{M}$)};
\STATE $i_{\min}=$ \textsc{GetMinDim(}\texttt{val\_errors}\textsc{)};
\STATE \texttt{selected\_dimensions}$[i_{\min}]=1$;
\STATE $\textbf{M}=\textbf{M}+\textbf{M}^{i_{\min}}$;
\ENDFOR
\STATE {\textbf{return}} {\texttt{selected\_dimensions}}
\end{algorithmic}%
\end{algorithm}

The procedure \textsc{GetValidationErrors} returns the validation errors for all
dimensions that have not yet been selected and contributes most to the overall
runtime. For each such dimension $j$, it computes a matrix $\widehat{\mathbf{M}}
= \mathbf{M} + \mathbf{M}^j$ containing all pairwise distances with the
distances of dimension $j$ being `added on the fly' to the distances that
correspond to the previously selected dimensions. This intermediate training set
is then used to compute the cross-validation error for the currently selected
set of dimensions. It turns out that this procedure and the overall workflow is
particularly well-suited for a massively-parallel implementation. Basically, one
can parallelise the search over all dimensions that have not yet been selected
as well as over the computations of the induced cross-validation errors. By
using a standard GPU device, one can reduce the runtime by a factor of up to 150
compared to single-core CPU implementation, hence, reducing the practical
runtime needed from hours to minutes only. We refer to \cite{GiesekePOI2014} for
the technical details and an experimental analysis of the runtimes for typical
astronomical data sets.

\revcolor
\section{Obtaining code and data}
\label{sec:code_and_data}
We want to make the results presented in this paper as reproducible as possible,
so we are releasing the code and the data obtained from SDSS. The code for the
GPU implementation of the nearest neighbours search is
available at GitHub: \url{https://github.com/gieseke/speedynn}.
The scripts and data for for reproducing the main results of this paper can be
found at \url{http://image.diku.dk/kstensbo/papers/1606.01/}.
The page contains a step-by-step guide to setting up the software and recreating
the main results presented in this paper.


\section{Results from feature selection}

\begin{figure*}
    \centering
    \includegraphics[width=.9\textwidth]{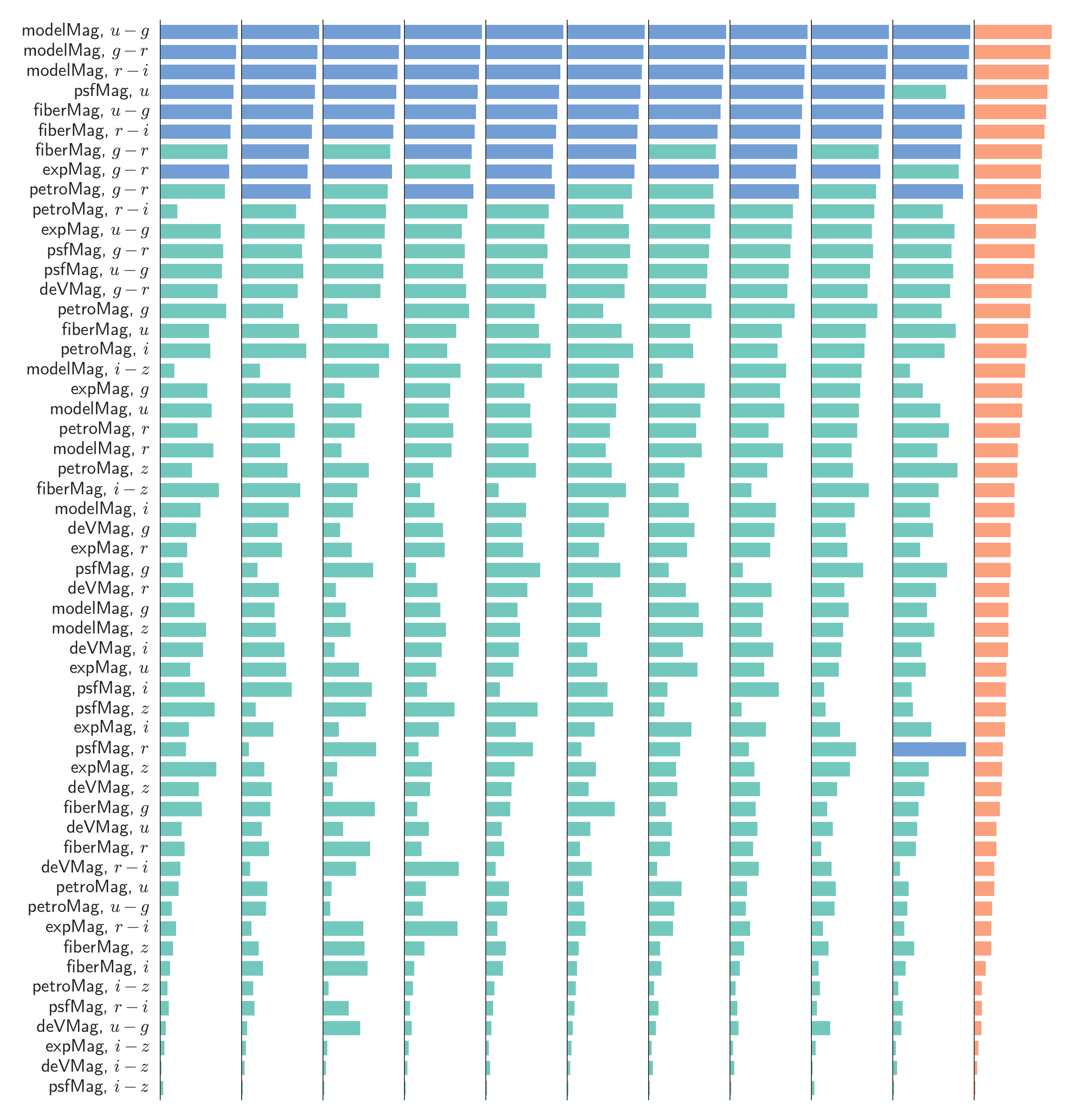}
    \caption{Ranking of features for \ssfr estimation according the feature
        selection in experiment 2.  To the left are the feature names, while the
        rightmost column shows the median rank of each feature across all CV
        folds. Each of the other columns shows the feature ranking in a
        particular CV fold. The larger the bar for a certain feature, the more
        important the feature was. Blue bars show features that were picked out
        during the feature selection as the most informative in a particular CV
        fold. Because of the differences in the data used in each CV fold the
        exact features picked out as important, as well as the number of chosen
        features per fold, will vary.  The number of chosen features vary
        between 7 and 9 with a median of 8.
    }
    \label{fig:full_feature_ranking_ssfr}
\end{figure*}

Figure~\ref{fig:full_feature_ranking_ssfr} shows the full feature ranking for the
\ssfr estimation done in experiment 2.

\begin{figure*}
    \centering
    \includegraphics[width=.9\textwidth]{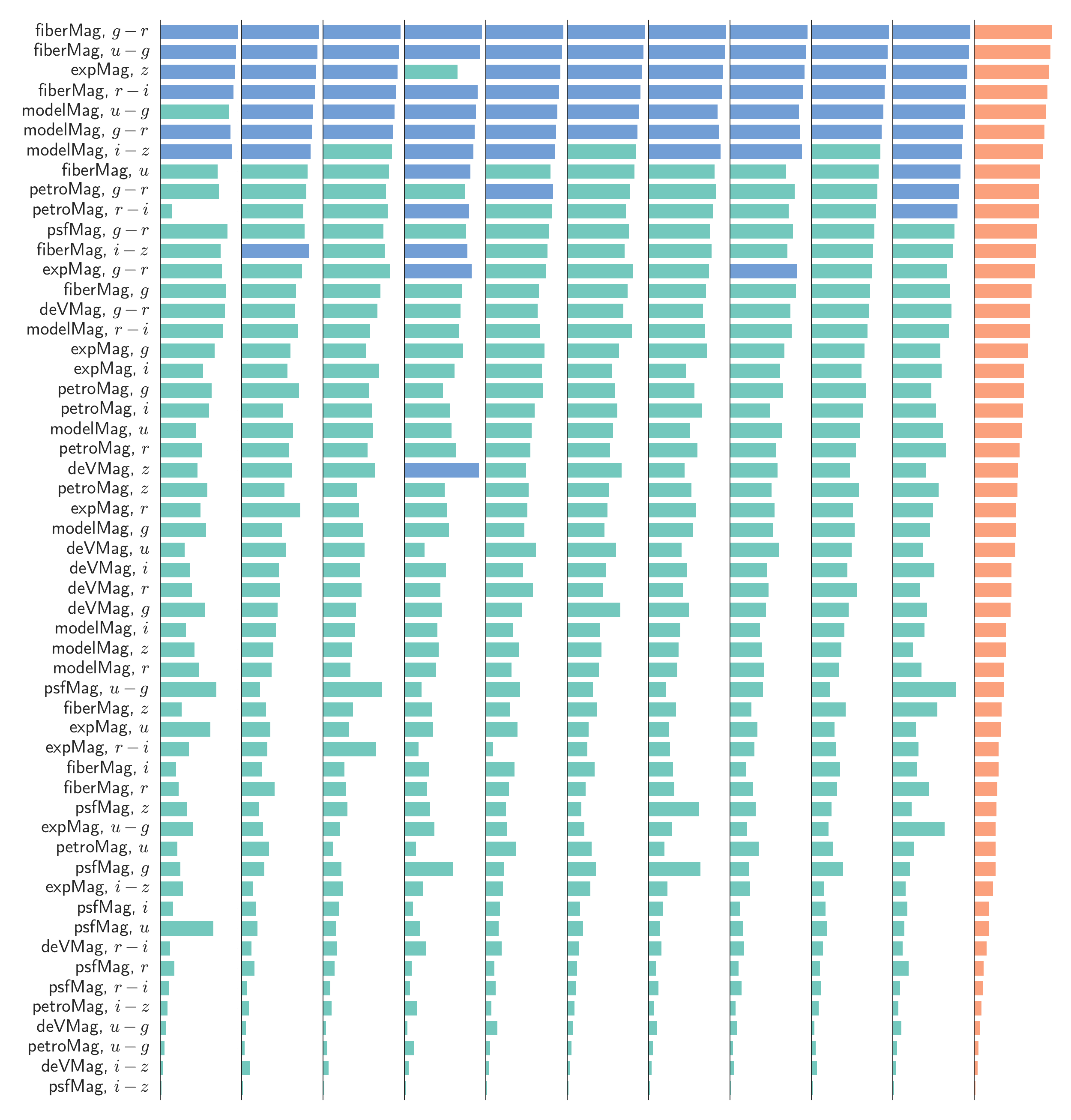}
    \caption{Ranking of features for \pz estimation according the feature
        selection in experiment 2.  To the left are the feature names, while the
        rightmost column shows the median rank of each feature across all CV
        folds. Each of the other columns shows the feature ranking in a
        particular CV fold. The larger the bar for a certain feature, the more
        important the feature was. Blue bars show features that were picked out
        during the feature selection as the most informative in a particular CV
        fold. Because of the differences in the data used in each CV fold the
        exact features picked out as important, as well as the number of chosen
        features per fold, will vary.  The number of chosen features vary
        between 6 and 11 with a median of 8.
    }
    \label{fig:full_feature_ranking_z}
\end{figure*}

Figure~\ref{fig:full_feature_ranking_z} shows the full feature ranking for the
\pz estimation done in experiment 2.

\section{Residual plots}
\label{app:residuals}

\begin{figure*}
    \centering
    \subfloat[Template-based model, small subset.\label{fig:xres:template_subset}]{
        \includegraphics[width=0.33\textwidth]{brinchmann/template_ssfr_residuals}
    }
    \subfloat[Experiment 1.\label{fig:xres:subset_ssfr}]{
        \includegraphics[width=0.33\textwidth]{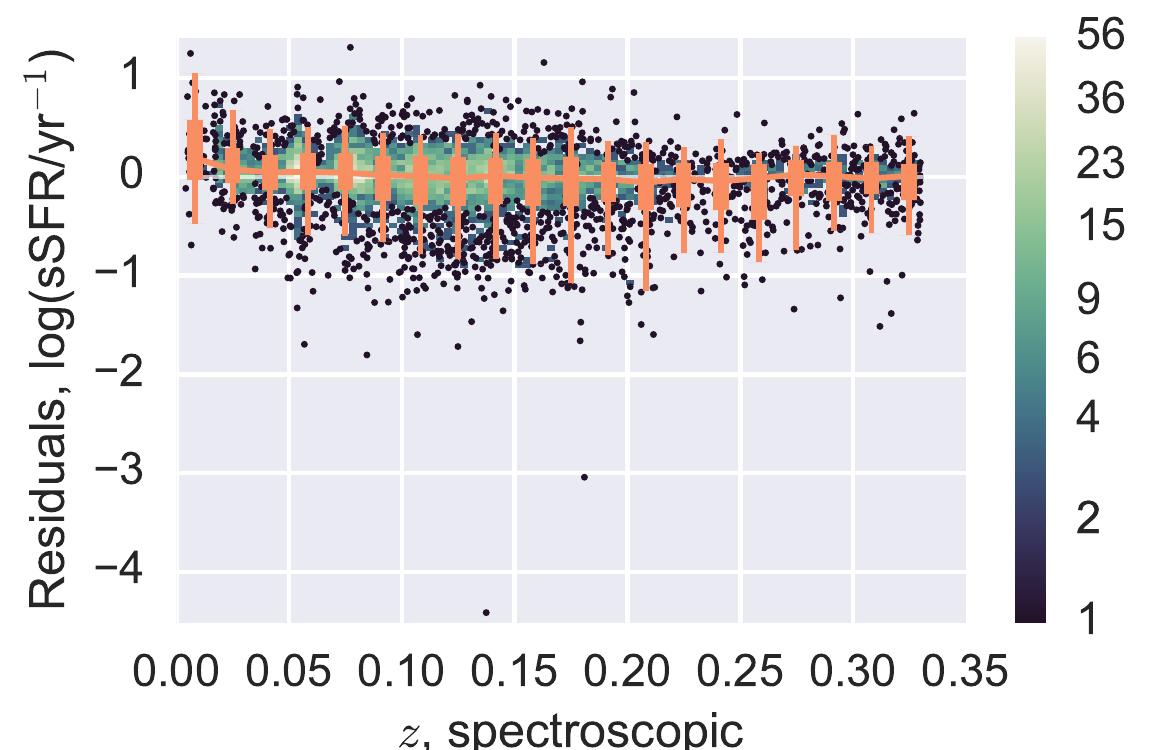}
    }
    %
    \subfloat[Experiment 3.\label{fig:xres:bigdata_ssfr}]{
        \includegraphics[width=0.33\textwidth]{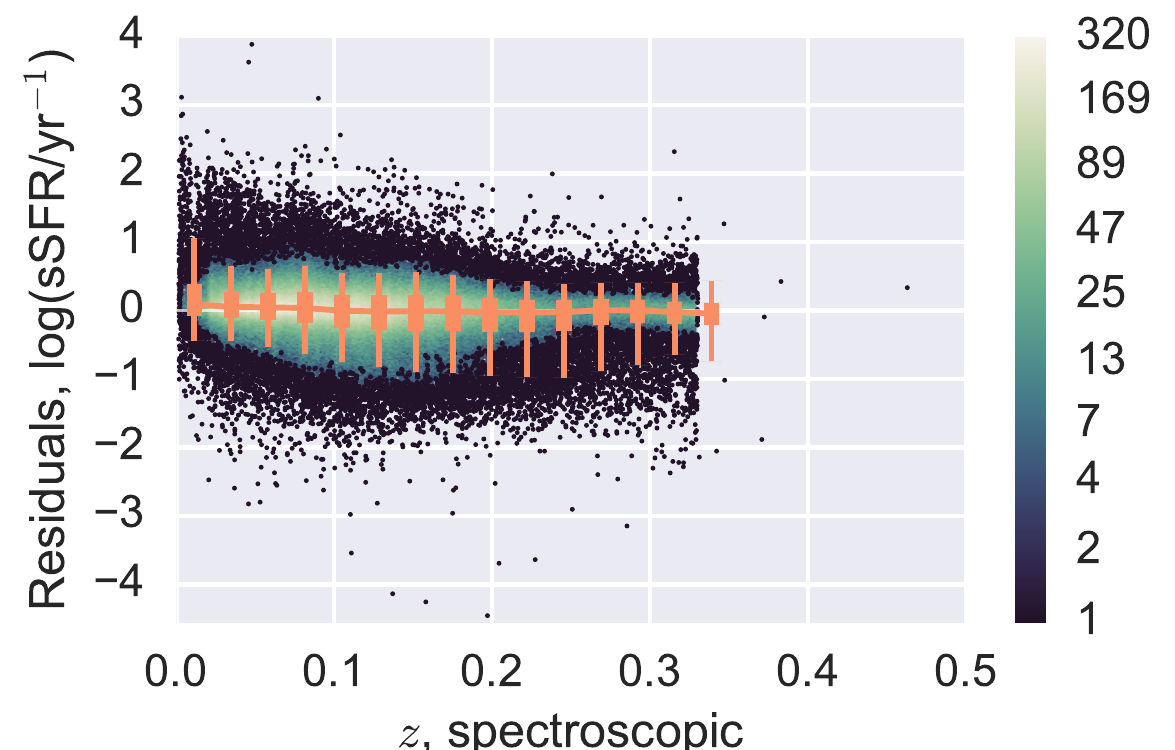}
    }
    \caption{\ssfr residuals as function of redshift for the two galaxy samples
        used in the experiments.  The colour coding of the distributions
        indicates the amount of galaxies in each bin.  The orange line shows the
        running median of the underlying distribution, the thick bars span the
        15.87th through the 84.13th percentile ($\pm 1\sigma$), and the thin
        bars span the 2.28th through the 97.72th percentile ($\pm 2\sigma$).
}
    \label{fig:extra_residuals_ssfr}
\end{figure*}

Residual plots for \ssfr experiment~1 and 3 can be seen in
Fig.~\ref{fig:extra_residuals_ssfr} together with residuals of the
template-based model, for comparison.

\begin{figure*}
    \centering
    \subfloat[SDSS \pz, small subset.\label{fig:xres:sdss_subset}]{
        \includegraphics[width=0.45\textwidth]{sdss/template_redshift_residuals}
    }
    \subfloat[SDSS \pz, large subset.\label{fig:xres:sdss_bigdata}]{
        \includegraphics[width=0.45\textwidth]{sdss/bigdata_redshift_residuals}
    }
    \\
    \subfloat[Experiment 1.\label{fig:xres:subset}]{
        \includegraphics[width=0.45\textwidth]{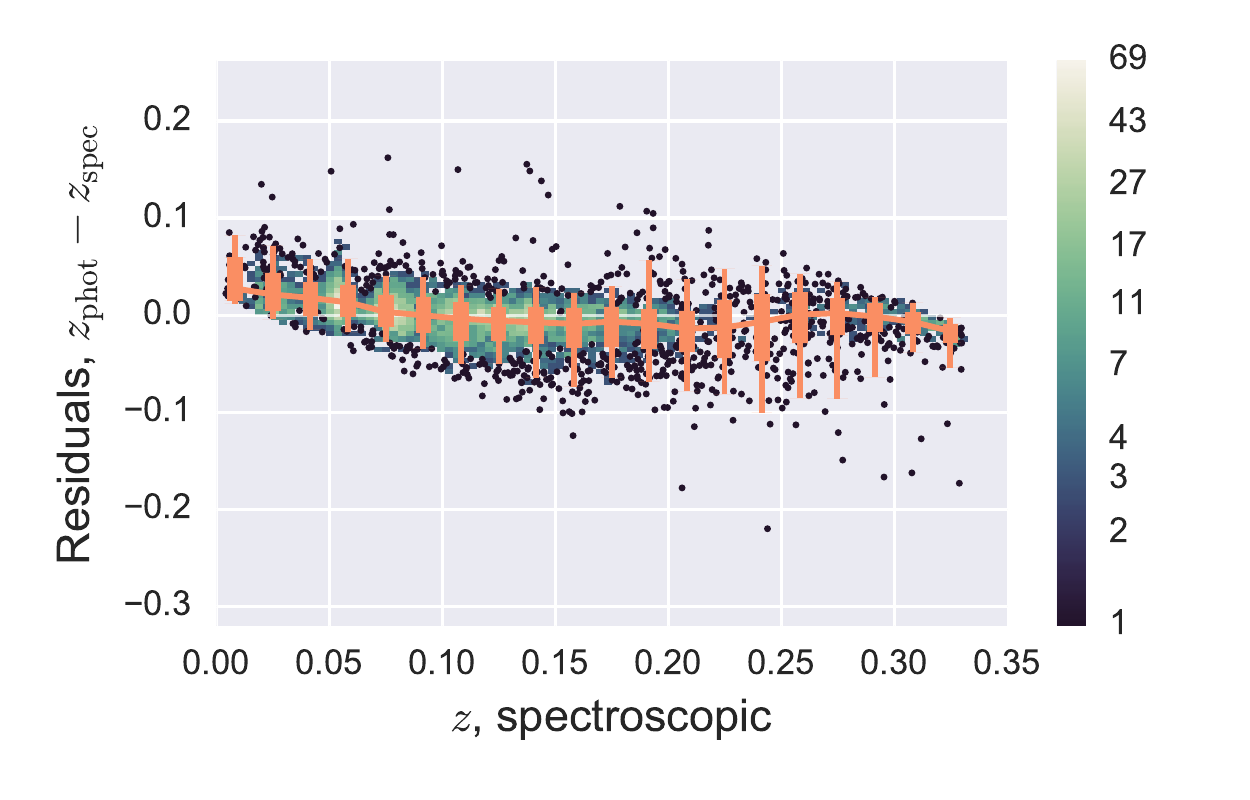}
    }
    \subfloat[Experiment 3.\label{fig:xres:bigdata}]{
        \includegraphics[width=0.45\textwidth]{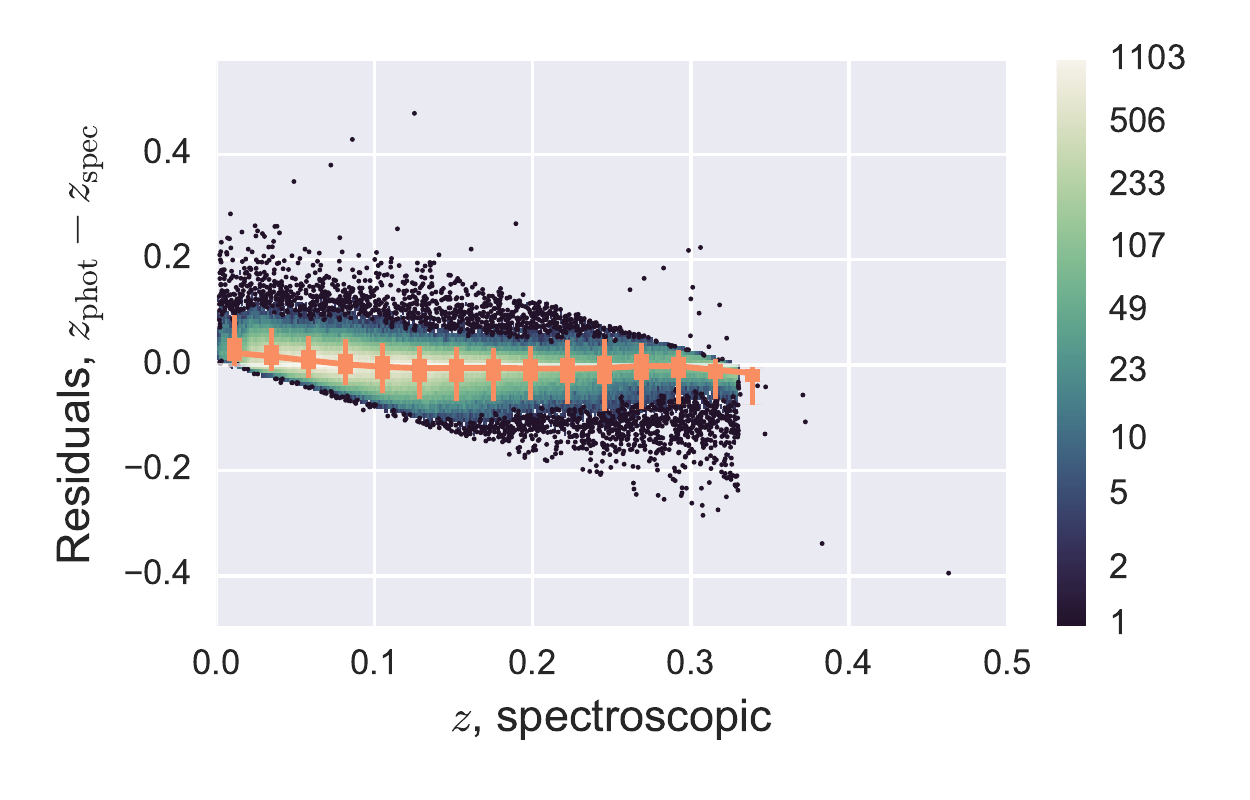}
    }
    \caption{Redshift residuals as function of redshift for the two galaxy
        samples used in the experiments.  The colour coding of the distributions
        indicates the amount of galaxies in each bin.  The orange line shows the
        running median of the underlying distribution, the thick bars span the
        15.87th through the 84.13th percentile ($\pm 1\sigma$), and the thin
        bars span the 2.28th through the 97.72th percentile ($\pm 2\sigma$).
        The sharp slopes seen in \protect\subref{fig:res:subset} and
        \protect\subref{fig:res:bigdata} are a consequence of the training set
        containing only few galaxies with $z \gtrsim 0.33$. As the \knn method
        is not well suited for extrapolation, only few galaxies will have an
    estimated \pz$\gtrsim 0.33$.  }
    \label{fig:extra_residuals_z}
\end{figure*}

Residual plots for \pzs experiment~1 and 3 can be seen in
Fig.~\ref{fig:extra_residuals_z} together with residuals of the SDSS
method for the same datasets, for comparison.

\color{black}


\bsp	
\label{lastpage}
\end{document}